\documentclass[12pt]{article}
\usepackage{amssymb}
\usepackage{epsf}
\usepackage{epsfig}
\usepackage{wasysym}
\usepackage{cite}
\usepackage{axodraw}
\newcommand{\lsim}{
\mathrel{\hbox{\rlap{\hbox{\lower4pt\hbox{$\sim$}}}\hbox{$<$}}}}

\newcommand{\gsim}{
\mathrel{\hbox{\rlap{\hbox{\lower4pt\hbox{$\sim$}}}\hbox{$>$}}}}

\newcommand{\comment}[1]{} 

\newcommand{\gev}{\, {\rm GeV}}
\newcommand{\mev}{\, {\rm MeV}}
\def\kpn{K^+\rightarrow\pi^+\nu\bar\nu}
\def\klpn{K_{ L}\rightarrow\pi^0\nu\bar\nu}

\begin{document}
\begin{titlepage}
\vspace*{-0.5truecm}

\begin{flushright}
TUM-HEP-658/07\\
UAB-FT/624\\
hep-ph/yymmnnn
\end{flushright}

\vspace*{0.3truecm}

\begin{center}
\boldmath
{\Large{\bf Flavor Changing Neutral Current Effects\\
\vspace*{0.3truecm}
and CP Violation in the Minimal 3-3-1 Model}}
\unboldmath
\end{center}

\vspace{0.9truecm}

\begin{center}
{\bf Christoph Promberger${}^a$, 
Sebastian Schatt${}^a$ and Felix Schwab${}^{b}$}
 
\vspace{0.5truecm}

${}^a$ {\sl Physik Department, Technische Universit\"at M\"unchen,
D-85748 Garching, Germany}\\

\vspace{0.2truecm}

${}^b$ {\sl Departament de F\'{\i}sica Te\`orica, IFAE, UAB, E-08193 Bellaterra,
 Barcelona, Spain}

\end{center}

\vspace{0.6cm}
\begin{abstract}
\vspace{0.2cm}\noindent
We investigate in detail the flavor structure of the minimal 331 model and its implications for several flavor changing neutral current (FCNC) 
processes. In this model, where the weak $SU(2)_L$ gauge group of the Standard Model is extended to a $SU(3)_L$, the by far dominant new
 contributions come 
from an additional neutral $Z'$ gauge boson, that can transmit FCNCs at tree-level. At the same time, electroweak precision observables receive
new contributions only at the loop level and do not constrain the model very strongly. In our analysis, we take into account new CP violating
effects that have been neglected in earlier analyses, and account for a general flavor structure without reference to a certain 
parameterization of the new mixing matrix. We begin by studying the bounds obtained from quantities such as $\Delta M_K$, $\epsilon_K$, 
$\Delta M_{d/s}$ as well as $\sin 2 \beta|_{J/\psi K_S}$, and go on to explore the implications for several clean rare decay channels, namely 
the decays $\kpn$, $\klpn$, $B_{d/s}\to \mu^+\mu^-$ and $K_L \to \pi^0 l^+ l^-$. We find sizeable effects in all these decays, but the most interesting quantity turns out to 
be the $B_s^0 - \bar B_s^0$ mixing phase $\beta_s$, as measured in the mixing induced CP asymmetry of $B_s^0 \to J/\psi \phi$, which can be 
large. In general, we find effects in purely hadronic channels to be larger than in (semi-)leptonic ones, due to a suppression of the
$Z'$-lepton couplings.
\end{abstract}

\vspace*{0.5truecm}
\vfill
\noindent
%\today

%
%
%
\end{titlepage}

\thispagestyle{empty}
\vbox{}
\newpage

\setcounter{page}{1}

\section{Introduction}\label{sec:intro}

The Standard Model of Particle Physics (SM) describes at present most of the observed phenomena in nature, with the exception of a 
consistent inclusion of gravitational effects. Still, there are several open questions remaining in this model, concerning, among others,
the matter of electroweak symmetry breaking, as well as the explicit particle content of the model, where there are three generations for both 
quarks and leptons. This latter question can be answered in the context of the 331 models \cite{Frampton:1992wt, Pisano:1991ee}, where anomaly 
cancellation and the asymptotic freedom of QCD require the number of generations to be precisely three. In order to do so, the $SU(2)_L$ 
doublet of the weak interactions is extended to a triplet with additional heavy quarks, and, additionally, the third generation transforms as an 
anti-triplet under the $SU(3)_L$. 

In the breaking process of this new, enlarged gauge group to the SM and, subsequently, its electromagnetic $U(1)_{em}$, additional gauge bosons are encountered, among 
these a neutral $Z'$ boson, which is naturally heavier than the SM gauge bosons, since its mass arises from the larger VEV that breaks the 
$SU(3)_L$ at a high scale. Similarly, there are heavy charged and doubly charged gauge bosons, as well as additional heavy, exotically charged (in the minimal 331 model)
quarks that constitute the third member of the $SU(3)_L$ triplet. In the leptonic sector these third triplet members are just given by the charge conjugated counterpart
of the charged SM lepton.

While the charged gauge bosons can appear for low energy processes involving quarks only at loop level, since they always couple also the heavy quark, the
neutral $Z'$ can transmit flavor changing neutral currents (FCNC) at tree level. Therefore, these processes can place 
rather stringent bounds on the mass of this heavy gauge boson, and there have been several analyses of certain FCNC observables in the 
literature \cite{Liu:1993gy,GomezDumm:1994tz,Rodriguez:2004mw}. In addition, the FCNC processes involving down type quarks are also affected
by the unitary quark mixing matrix used do diagonalize the down type Yukawa coupling, while those involving up type quarks appear with the 
corresponding up type mixing matrix. Thus, there is the possibility of new CP violating phases, which have, however, been neglected in 
all previous analyses of this type (see, on the other hand, \cite{Langacker:2000ju}, where the most general type of $Z'$ coupling is analyzed in a model independent manner). 

Also, it has been repeatedly pointed out in the literature \cite{Liu:1993gy,GomezDumm:1994tz,Rodriguez:2004mw}, that the most stringent FCNC constraints 
arise from parameters involving flavor mixing, in 
particular the mass differences in the neutral $K$ and $B$ meson systems, and the new measurement of $\Delta M_s$, the mass difference in the $B^0_s$ system 
is expected to have a significant impact here. In view of these two points we find it interesting to reanalyze in a 
complete manner the most important FCNC observables within the minimal 331 model, where we include also an analysis of several rare decay processes,
which have not been analyzed before. We would also like to point out, that FCNCs, which can provide lower bounds on the 
$Z'$ mass, are complementary to the corresponding {\em upper} bounds stemming from the fact that the model produces a Landau pole above a certain scale. 
However, these lower bounds are always obscured by some lack of knowledge of the mixing matrix elements.
Therefore, we will pursue in our analysis a route that is somewhat complementary to a standard FCNC analysis: We will 
not attempt to place lower bounds on the $Z'$ mass, but rather set its mass at several fixed values and will try to gain some information on the 
structure of the appearing quark mixing matrix. In addition, we will investigate the implications of the bounds obtained from well-measured observables
such as $\Delta M_K$, $\varepsilon_K$, $\Delta M_{d/s}$ and $\sin 2 \beta$ for several clean rare decays, where we can give upper bounds on the corresponding branching fractions
depending on the $Z'$ mass. Let us finally point out that the study of FCNC processes in these models
are particularly interesting as they occur at tree level, while the usual electroweak precision (EWP) observables, that 
strongly constrain most beyond SM models, occur only at the loop level, which actually makes the bounds from FCNC processes 
more stringent than those from EWP measurements. The most recent study of electroweak precision observables can be found in \cite{Long:1999yv}. 

Our paper is organized as follows: In Section \ref{sec:model}, we introduce the minimal 331 model, thereby also setting our conventions.
In addition, we give the FCNC vertices and a convenient parameterization of the corresponding quark mixing matrix in order to reduce the 
number of parameters appearing. Next, in Section \ref{sec:obs} we give the additional $Z'$ contributions to several observables, which we 
evaluate 
numerically in Sect. \ref{sec:numerics}. Among these observables, there are the mass differences in the neutral meson systems, as noted above, 
as well as several CP violating quantities, from which some information on the phase structure of the model can be obtained. During this numerical analysis, we
compare our work several times to a recent, similar analysis of the Little Higgs model with T-parity (LHT) performed in \cite{Blanke:2006eb}, since both models share 
the feature of introducing new CP violating phases while keeping the operator basis the same as in the SM.
Finally, Section \ref{sec:conclusions} contains our conclusions.

\section{The Minimal 331 Model}\label{sec:model}
Let us begin by introducing the particle content of the minimal 331 model. Many details of this model have been
first worked out in \cite{Ng:1992st}, to which we refer the reader for some more information.
The model consists of a gauge group $SU(3)_C \times SU(3)_L \times U(1)$, which is broken down in 
two steps:
\begin{equation}
SU(3)_C \times SU(3)_L \times U(1)_X \stackrel{v_{\sigma}}{\Rightarrow} SU(3)_C \times SU(2)_L \times U(1)_Y
\stackrel{v_{\eta},v_{\rho}}{\Rightarrow} U(1)_{em}
\end{equation}
Here, in contrast to the SM, two Higgs multiplets are required for the breaking of $SU(2)\times U(1)$ in order to
give masses to all quarks\footnote{Another Higgs, in a 6 representation is required for the lepton masses, but we will ignore it in the following, since it plays no role 
in our analysis.}. The additional VEV $v_{\sigma}$ is much larger than the two others. The charge assignment for the Higgs multiplets is as follows:
\begin{equation}
\sigma = \frac{1}{\sqrt{2}}\pmatrix{\sigma_1^{++} \cr \sigma_2^+ \cr v_{\sigma}+\xi_{\sigma}+i \zeta_{\sigma}} :(1,3,1) ,\quad
\            
\rho = \frac{1}{\sqrt{2}}\pmatrix{\rho_1^+ \cr v_{\rho}+\xi_{\rho} +i \zeta_{\rho} \cr \rho_2^-} : (1,3,0), 
\end{equation}
\begin{equation}
\eta = \frac{1}{\sqrt{2}}\pmatrix{v_{\eta}+\xi_{\eta}+i~\zeta_{\eta}  \cr \eta_1^- \cr \eta_2^{--} \cr} : (1,3,-1) 
\end{equation}
where the $\xi_i$ and $\zeta_i$ denote the real (scalar) and imaginary (pseudoscalar) fluctuations around the appropriate VEVs.
In analogy, the fermion content of the minimal model is given by
\begin{eqnarray}
\psi_{1,2,3} = \pmatrix {e\cr -\nu_e\cr e^c\cr} \: , \pmatrix {\mu \cr -\nu_\mu
   \cr \mu^c \cr} \: , \pmatrix {\tau \cr -\nu_\tau \cr \tau^c \cr}
     \qquad &\mathbin:& \qquad (1, \: 3^\ast, \: 0) \ , \\
Q_{1,2} = \pmatrix {u\cr d\cr D\cr} \: , \pmatrix {c\cr s\cr S\cr}
     \qquad &\mathbin:& \qquad (3, \: 3, \: -\textstyle\frac{1}{3} ) \ , \\
Q_3 = \pmatrix {b\cr -t\cr T\cr} \qquad &\mathbin:& \qquad (3, \: 3^\ast, \:
     \textstyle\frac{2}{3} ) \ ,  \\
d_R, \: s_R, \: b_R \qquad &\mathbin:& \qquad \ -\textstyle\frac{1}{3} \ ,  \\
u_R, \: c_R, \: t_R \qquad &\mathbin:& \qquad \textstyle\frac{2}{3} \ , \\
D_R, \: S_R \qquad &\mathbin:&  \qquad \ -\textstyle\frac{4}{3} \ ,  \\
T_R \qquad &\mathbin:& \qquad  \textstyle \frac{5}{3} \,
\end{eqnarray}
where the numbers in brackets correspond to the $SU(3)_C$, $SU(3)_L$ and $U(1)_X$ quantum numbers. For the right-handed fields, we give only the $U(1)$ number. From 
these, the electric charge can be obtained by 
\begin{equation}
Q=T_3+\sqrt 3 T_8 + X
\end{equation}
in our normalization of the charge $X$. In order to cancel anomalies, one generation of quarks has to transform as a $3^*$ under 
the $SU(3)_L$, and we choose this to be the third generation, but the explicit distinction only makes a difference once a specific 
structure of the mixing matrix is assumed. The factor $\sqrt 3$ can, in principle, be replaced by any number, thereby distinguishing
the different 331 models. Setting $\beta=-1/\sqrt{3}$, for example, requires a different fermion structure, and with it the introduction
of right-handed neutrinos \cite{Foot:1994ym}. 
This 331 model with right-handed neutrinos has also been under quite some discussion \cite{Montero:1992jk,Long:1999ij,Gutierrez:2004sb}, while analyses of models 
with general or
at least various different values of $\beta$ have been performed in \cite{Diaz:2004fs}. Also, there have been slight modifications added 
to the leptonic sector
in some models in order to generate neutrino masses \cite{Okamoto:1999cf,Kitabayashi:2000nq,Tully:2000kk,Montero:2001ts,Cortez:2005cp}, as well as supersymmetric versions 
of the model \cite{Duong:1993zn,Montero:2000ng,Montero:2004uy}. 

Let us next briefly summarize the gauge boson content of the model. The physical sector consists of three neutral gauge bosons, 
$A,Z$ and $Z'$, which arise as mass eigenstates from the diagonalization of the gauge boson mass matrix and are composed of the gauge eigenstates as
\begin{eqnarray} 
\label{Zmix} Z &=& + \cos \theta_W \, W_3 - \sin \theta_W \left( \sqrt{3} \tan \theta_W \,
       W_8 + \sqrt{1\!-\!3 \tan^2 \theta_W} \, B \right) \ ,  \\
\label{Zpmix} Z' &=& - \sqrt{1\!-\!3 \tan^2\theta_W} \, W_8 + \sqrt{3} \tan
      \theta_W \, B \ , \\
\label{Amix} A &=& + \sin \theta_W \, W_3 + \cos \theta_W \left( \sqrt{3} \tan \theta_W
         \, W_8 + \sqrt{1\!-\!3 \tan^2\theta_W} \, B \right)
\end{eqnarray}
In these formulae, the ratio between the $U(1)_X$ coupling $g_X$ and the $SU(3)_L$ coupling $g$ has already been expressed through the Weinberg-angle $\theta_W$:
\begin{equation} \label{WW}
\frac{g_X^2}{g^2} = \frac{6 \, \sin^2 \theta_W}{1 \!-\!4 \: \sin^2 \theta_W} \,.
\end{equation} 
In addition, there are the SM-like $W^{\pm}$ bosons, as well as another singly charged $Y^{\pm}$ boson, which transmits transitions from the second to third 
element of the triplets and a doubly charged bilepton $Y^{++}$, which transmits transitions from the first to the third element. We will mainly be concerned with the neutral 
sector in the following, and the corresponding masses are
\begin{eqnarray}
M^2_Z &=& \frac{1}{4} \: \frac{g^2}{\cos^2 \theta_W} \: (v_{\rho}^2+v_{\eta}^2) \,  \\
M^2_{Z'} &=& \frac{1}{3} \: g^2  \left( \frac{\cos^2 \theta_W}{1-4 \sin^2 \theta_W} v_{\sigma}^2 {} \right. \nonumber\\  
& & {}\left. \quad +\frac{1-4 \sin^2 \theta_W}{4 \cos^2 \theta_W } v_{\rho}^2 + \frac{(1+2 \sin^2 \theta_W)^2}{4 \cos^2 \theta_W (1-4 \sin^2 \theta_W)}  v_{\eta}^2 \right) \, \\
M^2_A &=& 0
\end{eqnarray}
which leaves indeed one massless photon, a $Z$ of the order of the weak scale as well as a heavier $Z'$. In principle, there can also be mixing between the
$Z$ and the $Z'$, but it is constrained to be small, see, e.g. \cite{Liu:1993fw}.
Finally, the scalar sector of this model has been analyzed in \cite{Anh:2000bs,Diaz:2003dk}, with the result that there is one light neutral Higgs, corresponding 
to the SM Higgs, 
three additional neutral heavy Higgs Fields as well as two singly charged and one doubly charged Higgs. In principle, these Higgs Fields should also 
transmit FCNCs, but these are suppressed by small Yukawa couplings of the external quarks and leptons in all processes we are studying.
Therefore, we will focus on the effects of the additional $Z'$, since these are expected to dominate and refer the reader to 
Refs. \cite{Ng:1992st,Liu:1993gy,Liu:1994rx,GomezDumm:1994tz,Diaz:2004fs}
for a more detailed analysis of the Yukawa coupling terms. Note also, that the relation (\ref{WW}) between
the coupling constants imposes additional constraints on the symmetry breaking scale $v_{\sigma}$ (and, in analogy, on the $Z'$ mass), in order to avoid the 
Landau pole that arises if $\sin^2 \theta_W=1/4$.
A careful analysis \cite{LP} shows that this scale can be several TeV. To be explicit, we take $5 \mathrm{TeV}$ as an upper bound, which is close to the number given,
for the case when exotically charged quarks are included, as we are doing here. 

\vspace{0.5cm}
The fact that the third quark family transforms differently under the $SU(3)_L$ leads to a flavor dependent $Z'$ coupling, as shown in Table \ref{TABLEFermions}, where 
we have collected the  neutral quark - gauge boson vertices in the weak eigenstate basis, writing $s_W \equiv \sin \theta_W$ and $c_W \equiv \cos \theta_W.$ 
In addition, we give also the coupling of the $Z'$ to leptons, which will also be 
required later on. This table is inspired by the similar table given in \cite{Perez:2004jc}, which is, however formulated in terms of vector and axial vector couplings.
The complete Lagrangian for the neutral currents, given in terms of these couplings, then reads: 
\begin{eqnarray} \label{LFergesgamma}
{\cal L}_{\rm Fermion}^{\rm NC}= & & i e \sum_f Q_f (\overline{f} \gamma_\mu f) A^\mu \nonumber\\
& &+ i \sum_f \left( \overline{f} \gamma_\mu (g_{l.h.}^{f Z} \gamma_L + g_{r.h.}^{f Z} \gamma_R) f Z^\mu + \overline{f} \gamma_\mu (g_{l.h.}^{f Z'} \gamma_L + g_{r.h.}^{f Z'} \gamma_R) f {Z'}^\mu \right) \,,
\end{eqnarray}
with $\gamma_{L/R}=\frac{1}{2}(1\mp\gamma_5)$.
Note that the lepton coupling is suppressed by a factor of $\sqrt{1-4s^2_W}$, which enhances the quark vertices, where it appears in the denominator. 
Therefore, the $Z'$ in the minimal 331 model has a somewhat leptophobic nature, which will become apparent in our numerical analysis. 

The difference between the first two and the third generation induces FCNCs transmitted by the $Z'$ boson at tree level. 
The structure of these couplings arises, when the couplings of all quarks are collected into one universal neutral current, where unitary mixing transformations drop out, as 
in the case of the SM neutral current. If this is done, one additional term remains, containing only third generation quarks and describing the difference of the couplings 
between the third and the first two generations. 
Transforming these left-over terms to the mass eigenstate basis yields a flavor changing interaction of the form
\begin{equation}\label{FCNC}
\mathcal{L}_{FCNC}=(g^{b,Z'}_{l.h.}-g^{d,Z'}_{l.h.})[\overline{u}\gamma_\mu\gamma_L
U_L^\dagger\pmatrix{0&&\cr&0&\cr&&1}U_Lu+\overline{d}\gamma_\mu\gamma_L
\tilde V_L^\dagger\pmatrix{0&&\cr&0&\cr&&1}\tilde V_Ld] {Z'}^{\mu}\,.
\end{equation}
The matrices $U_L$ and $\tilde V_L$ diagonalize the up and down - type Yukawa couplings respectively and then obviously obey 
\begin{equation}
U_L^{\dagger} \tilde V_L = V_{CKM}.
\label{CKM} 
\end{equation}
%where the matrices $U_L$ and $\tilde V_L$ diagonalize the up and down - type Yukawa couplings respectively and then obviously obey \[ U_L^{\dagger} \tilde V_L^\phantom{\ast} = V_{CKM}. \] 
We have added the tilde to distinguish between the SM CKM matrix and the mixing matrix for the down type quarks and will omit the subscript $L$ in what follows.

Next, the charged current vertices in this basis are then
\begin{displaymath}\label{CCWvert}
J_{W^+}^\mu=\overline{u}\gamma^\mu\gamma_L U_L^{\dagger} \tilde V d  =\overline{u}\gamma^\mu\gamma_LV_{CKM}d
\end{displaymath}
\begin{eqnarray}\label{CCvert}
J_{Y^+}^\mu&=&\overline{d}\gamma^\mu\gamma_L \tilde V^\dagger
\pmatrix{1&0\cr0&1\cr0&0}D+\overline{T}\gamma^\mu\gamma_L\pmatrix{0&0&1}
U_Lu\nonumber\\
J_{Y^{++}}^\mu&=&\overline{u}\gamma^\mu\gamma_LU_L^\dagger
\pmatrix{1&0\cr0&1\cr0&0}D-\overline{T}\gamma^\mu\gamma_L\pmatrix{0&0&1}\tilde V d \, .
\label{eq:qcc}
\end{eqnarray}
The corresponding charged currents in the leptonic sector are given as Feynman Rules in the App. ~\ref{sec:FR}, where we also give the explicit expression of the Feynman 
Rules for the FCNC vertices. We follow \cite{Liu:1994rx}, in that we show these couplings in a basis in which the heavy $D$ and $S$ quarks are mass as well as gauge 
eigenstates. This fact explains the absence of an explicit mixing matrix for these heavy quarks. We have also combined them into a doublet, denoted simply as $D$ in the 
above formulae, and put the heavy $T$ into a separate singlet to simplify the notation.

In contrast, the left handed part of the neutral current coupling to the $Z$ boson is given by 
\begin{equation}
\mathcal{L}_Z = \frac{g}{\cos^2 \theta_W} (T_3- Q_f \sin^2 \theta_W) \bar q_L \gamma^{\mu} q_L Z_{\mu}\,, 
\end{equation}
as in the SM and does not discriminate between generations, so that these vertices remain flavor conserving. 
To find a sensible parameterization for the matrix $\tilde V$, we should first count the number of additional parameters that appear in this matrix. Looking at all 
the possible interaction terms, one finds that, after the phase transformations of the up and down-type quarks have been used to simplify the CKM matrix, there are
three more possible phases that arise from transformations in the $D,S,T$ quarks as seen in (\ref{CCvert}), which leaves one 
with 6 additional parameters, namely three mixing angles and three phases. However, from (\ref{FCNC}) it is obvious that only the $\tilde V_{3j}$ elements 
are required when calculating FCNCs, and it is possible to find a parameterization that further reduces the number of parameters appearing there. It reads
\begin{eqnarray}\label{eq:param}
\tilde V%&=&\pmatrix{v_{1d}&v_{1s}&v_{1b}\cr v_{2d}&v_{2s}&v_{2b}\cr
%v_{3d}&v_{3s}&v_{3b}} \\
&=&\pmatrix{\tilde V_{1d} & \tilde V_{1s} & \tilde V_{1b} \cr \tilde V_{2d} & \tilde V_{2s} & \tilde V_{2b} \cr \tilde V_{3d} & \tilde V_{3s} & \tilde V_{3b}}  \\ 
&=& \pmatrix{c_{12} c_{13} & s_{12} c_{23} e^{i \delta_3}-c_{12} s_{13} s_{23} e^{i(\delta_1- \delta_2)} & c_{12} c_{23} s_{13} e^{i \delta_1} + s_{12} s_{23} e^{i(\delta_2 + \delta_3)} \cr
-c_{13} s_{12} e^{-i \delta_3} & c_{12} c_{23} + s_{12} s_{13} s_{23} e^{i(\delta_1- \delta_2 - \delta_3)} & -s_{12} s_{13} c_{23} e^{i(\delta_1- \delta_3)} - c_{12} s_{23} e^{i \delta_2} \cr
-s_{13} e^{-i \delta_1} & -c_{13} s_{23} e^{-i \delta_2} & c_{13} c_{23} } \nonumber\ ,
\end{eqnarray}
where only two additional CP violating quantities $\delta_1$  and $\delta_2$ appear, that are responsible for the additional CP violating effects to be 
discussed below. Note, that these CP violating phases have been neglected in all previous analyses of FCNCs in 331 models. Note also, that the mixing 
angle $\theta_{12}$ does not appear in the relevant matrix elements. In choosing the parameterization of the matrix in such a way, one has to 
be careful to choose one that can actually be achieved by rotating the heavy $D,S$ and $T$ quarks, and a general unitary matrix with the correct 
number of parameters may not necessarily be allowed. However, we have checked that the parameterization (\ref{eq:param}) is. A similar parameterization, 
sharing several features but ignoring weak phases, can be found in \cite{Liu:1994rx}. 

Let us finally also comment on the corresponding vertices in the up-type sector of the model. In this case, there are no further phase transformations that can be performed,
so that the matrix $U_L$ can be just any arbitrary unitary matrix with, correspondingly, nine parameters, i.e. three angles and six phases, subject to the constraints from 
(\ref{CKM}). Additionally, the 
observables associated with $D$ mixing and decay are afflicted with rather large uncertainties coming from long distance QCD effects. Therefore, we will not
investigate these quantities any further in the course of this work.

\begin{table} 
\caption{\label{TABLEFermions} List of couplings for the neutral currents in the minimal $331$ model. In the corresponding Feynman Rules, an additional 
factor $i$ will appear. We abbreviate $s_W \equiv \sin \theta_W$ and $c_W \equiv \cos \theta_W.$ }
\begin{tabular}{llllll}
Fermion & $Q_f$ & $g^{f,Z}_{l.h.}$ & $g^{f,Z}_{r.h.}$ & $g^{f,Z'}_{l.h.}$ &
$g^{f,Z'}_{r.h.}$ \\
\hline 
$l^-$ & $-1$ & $-\frac{g (1-2s^2_W)}{2 c_W}$ & $\frac{g s_W^2}{c_W}$ &
$\frac{g \sqrt{1-4s^2_W}}{2 \sqrt{3}  c_W}$ &
$\frac{g \sqrt{1-4s^2_W}}{\sqrt{3} c_W}$ \\

$\nu_l$ & $0$ & $\frac{g}{2 c_W}$ & $0$ &
$\frac{g \sqrt{1-4s^2_W}}{2\sqrt{3} c_W}$ & $0$ \\

$u,c$ & $+\frac{2}{3}$ & $\frac{g (3-4s^2_W)}{6 c_W}$ & $-\frac{2 g s_W^2}{3 c_W}$ &$-\frac{g (1-2s^2_W)}{2\sqrt{3}c_W \sqrt{1-4s^2_W}}$ &
$\frac{2 g s^2_W}{\sqrt{3}c_W\sqrt{1-4s^2_W}}$ \\

$d,s$ & $-\frac{1}{3}$ & $ -\frac{g (3-2s^2_W)}{6 c_W}$ & $\frac{g s_W^2}{3 c_W}$
& $ -\frac{g (1-2 s_W^2)}{2\sqrt{3}c_W\sqrt{1-4s^2_W}}$ &
$-\frac{g s_W^2}{\sqrt{3}c_W\sqrt{1-4s^2_W}}$ \\

$D,S$ & $-\frac{4}{3}$ & $\frac{4 g s^2_W}{3 c_W}$ & $\frac{4 g s^2_W}{3 c_W}$ &
$\frac{g (1-5s^2_W)}{\sqrt{3}c_W\sqrt{1-4s^2_W}}$ &
$-\frac{4 g s_W^2}{\sqrt{3} c_W \sqrt{1-4s^2_W}}$\\

$b$ & $-\frac{1}{3}$ & $-\frac{g (3-2 s^2_W)}{6 c_W}$ & $\frac{g s_W^2}{3 c_W}$ &
$\frac{g}{2\sqrt{3}c_W \sqrt{1-4s^2_W}}$ &
$-\frac{g s^2_W}{\sqrt{3}c_W\sqrt{1-4s^2_W}}$ \\

$t$ & $+\frac{2}{3}$ & $\frac{g (3-4s^2_W)}{6 c_W}$ & $-\frac{2 g s_W^2}{3 c_W}$ & $\frac{g}{2\sqrt{3}c_W\sqrt{1-4s^2_W}}$ &
$\frac{2 g s^2_W}{\sqrt{3}c_W \sqrt{1-4s^2_W} }$ \\

$T$ & $+\frac{5}{3}$ & $-\frac{5 g s^2_W}{3 c_W}$ & $-\frac{5 g s^2_W}{3 c_W}$ &
$-\frac{g (1-6 s^2_W)}{\sqrt{3}c_W\sqrt{1-4s^2_W}}$ &
$\frac{5 g s_W^2}{\sqrt{3} c_W \sqrt{1-4s^2_W}}$\\
\end{tabular}
\end{table}

\section{Formulae for Observables}\label{sec:obs}
In this section, we will collect the theoretical expressions for all observables relevant to our analysis. In particular, we give the $Z'$ contributions that modify
the SM amplitudes. These will be investigated numerically in Sect \ref{sec:numerics}.

\subsection{Modifications in Meson Mixing Amplitudes}
We will first be concerned with observables related to $B_{d/s}^0 - \bar B_{d/s}^0$ and $K^0 - \bar K^0$ mixing. These are the mass differences $\Delta M_K$,
 $\Delta M_d$ and $\Delta M_s$, as well as the CP violating quantities $\epsilon_K$, $A^{mix}_{CP}(B_d^0 \to J/\psi K_S)$ and $A^{mix}_{CP}(B_s^0 \to J/\psi \phi)$. 
In all cases, we will concentrate on the contribution from the heavy $Z'$ bosons, while the heavier charged gauge bosons appear only at the one loop level. They can be 
probed,
for example, in the inclusive decay $b \to s \gamma$, where the tree level terms remain absent \cite{Agrawal:1995vp}, or similarly through decays such as 
$Z \to b \bar b$ \cite{Perez:2004jc,Gonzalez-Sprinberg:2005zd}. On the other hand, there are contributions to muon decay from these heavy charged gauge bosons. Since the 
coupling of these heavy bosons is exactly the same as the $W^\pm$ coupling to the leptons, this new piece 
can just be absorbed into a redefinition of the coupling constant as follows: $G_F = G_F^{\mu}/(1+(M_W/M_Y)^2)$, where $G_F^{\mu}$ is the coupling constant
measured in muon decay, while $G_F$ is the ``true'' coupling $G_F$, obeying $G_F/\sqrt{2}=g^2/(8 M_W^2)$, with $g$ the $SU(3)_L$ gauge coupling. 
To reduce the number of parameters appearing, we will assume that both the $Y^\pm$ and $Z'$ are given entirely by those contributions
stemming from the heaviest VEV, and express the $Y^\pm$ mass through $M_{Y^\pm}=3 (1-4 \sin^2 \theta_W)/(4 \cos^2 \theta_W) M_{Z'}$. 
This procedure leads, for example, to $G_F/G_F^{\mu}=0.92$ for $M_{Z'}=1 \mathrm{TeV}$. Note, that these effects appear only in 
the lepton sector, since here the third particle of the triplet is again a SM particle. In the quark sector, however, there are no new tree-level contributions
from the new charged gauge bosons, since these always couple to a heavy quark. Let us finally quote \cite{Ng:1992st}, where a lower bound of $M_{Y^\pm}>270 \gev$ is 
found from muon decay. Since, in our approximation of the $Y^\pm$ mass, this charged gauge boson is about 3 times lighter than the $Z'$, we shall also use 
$1 \: \mathrm{TeV}$ as a lower bound for $M_{Z'}$ in our analysis. A similar bound on $M_{Y^\pm}$ has been obtained from EWP tests in \cite{Long:1999yv}. 

From the FCNC Lagrangian and the neutral current couplings given above, we find the tree-level effective Hamiltonian for $\Delta F=2$ transitions, where $F=S$:
\begin{equation}\label{Heff}
H^{eff}_{\Delta S=2}= \frac{G_F}{\sqrt 2} \frac{1}{3} \frac{\cos^4 \theta_W}{1-4 \sin^2 \theta_W} \left( \frac{M_Z}{M_{Z'}} \right)^2 (\tilde V_{31} 
\tilde V_{32}^*)^2 (\bar s d)_{V-A} (\bar s d)_{V-A} \, ,
\end{equation}
while, in the $F=B$ case, the vertex factors are replaced by $\tilde V_{3q} \tilde V_{33}^*$ with $q=1,2$ for down and strange quarks, respectively. 
Since the $Z'$ induces FCNCs left-handedly, no new operators are generated, while, in general, there are obviously new sources of flavor and CP violation in
the Matrix $\tilde V$, so that the model does go beyond the usual minimal flavor violating (MFV) scenarios (see \cite{Buras:2003jf} for a review and a discussion of
the several definitions of MFV that are being used). 

Next, we need to take into account
the different nature of $B$ and $K$ mixings: While $B_{d/s}^0 - \bar B_{d/s}^0$ mixing proceeds through the absolute value of the corresponding matrix elements, 
$K^0 - \bar K^0$ mixing is described by the real part only (this distinction has been missed in the literature, note that this is not
even correct in the case of vanishing CP violation in $ \tilde V$ because of the phase in $V_{td}$). Therefore, we have
\begin{eqnarray}
\Delta M_K^{Z'} &=& \frac{G_F}{\sqrt 2} \frac{8}{9} \frac{\cos^4 \theta_W}{1-4 \sin^2 \theta_W} \left( \frac{M_Z}{M_{Z'}} \right)^2 
                     \mathrm{Re} [ (\tilde V_{31} \tilde V_{32}^*)^2] \hat B_{K}F_{B_K}^2 m_{K} \,, \\\label{DeltaMq}
\Delta M_{q}^{331} &=&\left| \Delta M_q^{SM} e^{-i 2 \beta}+ \frac{G_F}{\sqrt 2} \frac{8}{9} \frac{\cos^4 \theta_W}{1-4 \sin^2 \theta_W} 
\left( \frac{M_Z}{M_Z'} \right)^2 (\tilde V_{3q} \tilde V_{33}^*)^2 \hat B_{B_q}F_{B_q}^2 m_{B_q} \right| \,.
\end{eqnarray}
where we have given only the $Z'$ contribution in the case of $\Delta M_K$, but the complete expression containing the SM as well as the new contribution in 
the case of $\Delta M_{d/s}$.
The corresponding SM contributions are (see \cite{Buras:2005xt} for a review)
\begin{eqnarray}
\Delta M_K^{SM} &=& \frac{G_F^2}{6 \pi^2}  \hat B_{K}F_{B_K}^2 m_{K} M_W^2 \mathrm{Re} \left[ \eta_1 S_0(x_c) (V_{cs}^* V_{cd})^2+\eta_2 S_0(x_t) 
(V_{ts}^* V_{td})^2 + \right.\\ \nonumber && \left.2 \eta_3 V_{cs}^* V_{cd} V_{ts}^* V_{td} S_0(x_c,x_t)\right] \\
\Delta M_{q}^{SM} &=& \frac{G_F^2}{6 \pi^2} \eta_B  \hat B_{B_q}F_{B_q}^2 m_{B_q} M_W^2 S_0(x_t) |V_{tq}|^2\label{DeltaMqSM}
\end{eqnarray}
where, in the SM prediction, $\eta_1=1.32\pm0.32$, $\eta_2=0.57\pm0.01$, $\eta_3=0.47\pm0.05$  and $\eta_B=0.55 \pm 0.01$ are the NLO QCD corrections and the 
$S_0(x_i)$ are the leading order Inami Lim Functions that describe the charm and top box diagrams.

The contribution to the kaon CP violating parameter $\epsilon_K$ can also easily be calculated from the effective Hamiltonian (\ref{Heff}). It is  
\begin{equation}
\epsilon_K^{Z'}=\exp{(i \pi/4)}   \frac{G_F}{9} \frac{2~ M_K}{\Delta M_K} \frac{\cos^4 \theta_W}{1-4 \sin^2 \theta_W} \mathrm{Im} \left[ (\tilde V_{32}^* \tilde V_{31})^2 \right] \hat B_{K}F_{K}^2 \, ,
\end{equation}
where we use the experimental value for $\Delta M_K$ in our numerical analysis. Note that the new contributions to both $\epsilon_K$ as well as $\Delta M_K$ are simply
added to the SM contributions, i.e. there are no interference terms, while this is true in the case of $\Delta M_{d/s}$ only if the new contribution comes with the 
same phase as the SM contributions, as can be seen from (\ref{DeltaMq}). Let us also here give the SM expression, reading
\begin{equation}\label{epsilonKSM}
\epsilon_K^{SM}=e^{i \frac{\pi}{4}} \frac{G_F^2}{12\pi^2} \frac{M_K}{\sqrt{2}\Delta M_K} M_W2 [{\lambda_c^{\ast}}^2 \eta_1 S_0(x_c) + {\lambda_t^{\ast}}^2 \eta_2 S_0(x_t) + 2 \lambda_c^\ast \lambda_t^\ast \eta_3 S_0(x_c,x_t)] \hat B_{K}F_{K}^2 \,.
\end{equation}

Next, before we turn to the analysis of CP violating $B$ decay asymmetries, let us give the contributions that modify the $B_{d}^0 - \bar B_{d}^0$ mixing phase, 
which is equal to $2 \beta$ in the SM, where $\beta$ is one of the angles of the unitarity triangle. Including the additional contributions from the $Z'$, we find
\begin{eqnarray}\label{Phidcorrection}
\Phi_d^{331}&=&-\arg \left( M_{12}^{SM}+M_{12}^{Z'} \right) \\ \nonumber
&=&-\arg \left( \frac{G_F^2}{6 \pi^2} \eta_B  M_W^2 S_0(x_t) |V_{td}|^2 e^{-i 2 \beta}+ 
\frac{G_F}{\sqrt 2} \frac{8}{9}  \frac{\cos^4 \theta_W}{1-4 \sin^2 \theta_W} (\tilde V_{33}^* \tilde V_{31})^2 \left( \frac{M_Z}{M_{Z'}} \right)^2 \right)
%\\&\approx&- \left( 2  \beta + \frac{180 \tan (\arg (\tilde V_{33}^* \tilde V_{31})^2-2 \beta)}{\pi} \frac{|M_{12}^{Z'}|}{|M_{12}^{SM}|} \right)\, .
\end{eqnarray}
%In the last step, we have expanded, assuming that the absolute value of the new contribution is small. In our numerical analysis, we will, however, use the complete formula
%since we find that the new contributions are not small enough to warrant this kind of expansion.

In addition, there are also new contributions to decay amplitudes, in particular also to the amplitude of the decay $B \to J/\psi K_S$. In the 
SM, this decay proceeds through a tree diagram topology with no additional CP violating phase, so that the mixing induced CP asymmetry is 
given by $\sin 2\beta$. In a general model, $\beta$ is replaced by a value $\beta_{eff}$, which is given as 
$2 \beta_{eff}=\Phi_d+\Phi_{decay}$. Unfortunately, the $Z'$ couples also right-handedly to the charm quark pair, and we can therefore not simply add the
coefficient of the new tree diagram to the SM contribution. We have then estimated the projection onto the left-handed SM operator, which is entirely negligible,
and we therefore consider it a good approximation to omit these terms. Analogous modifications occur in the asymmetry of $B_s \to J/\psi \phi$, 
which in the SM is given by $\sin 2 \beta_s$ with $\beta_s=-2 ^{\circ}$. Including the new contribution, 
%Since we have only an additional tree diagram, whose coefficient can be absorbed into the entire 
%coefficient of the tree diagram like amplitude, the nice feature of the decay being dominated by just one amplitude remains, where in this
%case we have a weak phase
%\begin{equation}\label{decayphase}
%\theta \equiv \arg \left(  \frac{G_F}{\sqrt 2} V_{cs} V_{cb}^{*} + \frac{2 g^2 s_W^2}{3 \sqrt{1-4 s_W^2} M_{Z'}^2 }\tilde V_{33} \tilde V_{32}^* \right)
%\end{equation}
%and the mixing induced CP asymmetry now measures $\sin \Phi_d-2 \theta$. This contribution is the same for the decay $B_s \to J/\psi \phi$, 
%which in the SM is given by $\sin 2 \beta_s$ with $\beta_s=-2 ^{\circ}$. Including the new contribution, 
\begin{equation}
\Phi_s^{331}=-\arg \left( \frac{G_F^2}{6 \pi^2} \eta_B  M_W^2 S_0(x_t) |V_{ts}|^2 e^{-i 2 \beta_s}+ 
\frac{G_F}{\sqrt 2} \frac{8}{9}  \frac{\cos^4 \theta_W}{1-4 \sin^2 \theta_W} (\tilde V_{33}^* \tilde V_{32})^2 \left( \frac{M_Z}{M_{Z'}} \right)^2 \right) \,.
%5\phi_s \approx 2 \beta_s + \frac{180 \tan( \arg (\tilde V_{33}^* \tilde V_{32})^2-2 \beta_s)}{\pi} \frac{|M_{12}^{Z'}|}{|M_{12}^{SM}|} \,.
\end{equation}

We note, finally, that the observables discussed in this subsection are, in principle, sufficient to determine all the parameters appearing in our parameterization
of the mixing matrix (\ref{eq:param}). Also, the experimental situation for these observables will be summarized when we perform our numerical analysis in Section 
\ref{sec:numerics}.

\subsection{Modification in Rare Decay Amplitudes}

The observable quantities listed so far are all related to meson mixing, and have also all been measured (with the exception of $\beta_s$). 
Therefore, we will use them in the next section to 
constrain the parameter space of the model. Then, we will be interested in the implications of the bounds obtained in that 
analysis on several rare decay amplitudes. Most of the corresponding branching fractions have not yet been measured, but the measurements
will tell us quite a lot about the new physics contributions, since the theoretical expressions for these decays are extremely clean. The rare decays which 
we will study are $K^+ \to \pi^+  \nu \bar \nu$, $K_L \to \pi^0 \nu \bar \nu$, $B_{d/s} \to \mu^+ \mu^-$ and $K_L \to \pi^0 l^+ l^-$, where $l$ can be a 
muon or an electron.

Let us then begin this subsection with some general remarks: The rare decays in question are governed by electroweak- and photon-penguins as well as 
leptonic box diagram contributions. These are described in the Standard Model by the corresponding Inami Lim Functions $C_0(x_t)$, $D_0(x_t)$ and 
$B_0(x_t)$. In the expressions for decay amplitudes, these always appear in the gauge invariant combinations $X_0(x_t)$, $Y_0(x_t)$ and $Z_0(x_t)$ \cite{Buchalla:1990qz}, 
defined as:
\begin{equation}
C_0(x_t)-4B_0(x_t)=X_0(x_t)
\end{equation}
\begin{equation}
C_0(x_t)-B_0(x_t)=Y_0(x_t)
\end{equation}
\begin{equation}
C_0(x_t)+{1\over4}D_0(x_t)=Z_0(x_t).
\end{equation}
In models of minimal flavor violating type, the new contributions to decay amplitudes can often be absorbed into a universal redefinition of these 
functions. On the other hand, these functions will be process-dependent in models that go beyond minimal flavor violation, as explicitely discussed for 
the Littlest Higgs model with T-parity in \cite{Blanke:2006eb}.
We will see later that the situation is even slightly more complicated in the minimal 331 model. In the following, we will, whenever possible, give the 
appropriate redefinition of $X(x_t)$, $Y(x_t)$ and $Z(x_t)$ (the functions  without the subscript 0 always refer to the NLO Functions, while
those where it is included are only the LO ones) as
\begin{eqnarray}
X´_i(x_t)&=&X^{SM}(x_t)+\Delta X_i\,, \\
Y´_i(x_t)&=&Y^{SM}(x_t)+\Delta Y_i\,,\\\,
Z´_i(x_t)&=&Z^{SM}(x_t)+\Delta Z_i\,.
\end{eqnarray}

We begin with the cleanest rare decays, i.e. $K\to \pi \nu \bar \nu$, and $B_{d/s} \to \mu^+ \mu^-$. 
For the decay $K\to \pi \nu \bar \nu$ there exists a charged and a neutral counterpart, $\kpn$ and $\klpn$ \cite{Buras:2004uu}. 
Both decays are theoretically extremely clean, since the leading QCD matrix element can be extracted from 
the well measured tree-level decay $K^+\to\pi^0 e^+\nu$ and additional long-distance QCD effects are rather well under control \cite{Isidori:2005xm}.
The effective Hamiltonian consists of contributions from both charm and top-loops, and is then given by:
\begin{equation}
H^{\rm SM}_{\rm eff}={G_{\rm F} \over{\sqrt 2}}{\alpha\over 2\pi 
\sin^2\theta_W}
 \sum_{l=e,\mu,\tau}\left( V^{\ast}_{cs}V_{cd} X^l_{\rm NL}+
V^{\ast}_{ts}V_{td} X(x_t)\right)
 (\bar sd)_{V-A}(\bar\nu_l\nu_l)_{V-A} \,. \\
\end{equation}
Defining $\lambda_i=V^*_{is}V_{id}$ and collecting the charm contributions in $P_c(X)=0.41\pm0.05$\cite{Buras:2006gb,Isidori:2005xm}, the branching 
fraction for $\klpn$ and $\kpn$ can then be derived as
\begin{eqnarray}\label{bkpnn}
\mbox{BR}(K^+)&\equiv&\mbox{BR}(K^+\to\pi^+\nu\bar\nu) \\ \nonumber
&=&\kappa_+\cdot
\left[\left({\rm Im}\left(\frac{\lambda_t}{\lambda^5}X(x_t)\right)\right)^2+
\left({\rm Re}\left(\frac{\lambda_c}{\lambda}P_c(X)\right)+
{\rm Re}\left(\frac{\lambda_t}{\lambda^5}X(x_t)\right)\right)^2\right],
\end{eqnarray}
\begin{equation}\label{kapp}
\kappa_+=r_{K^+}\frac{3\alpha^2 \mbox{BR}(K^+\to\pi^0 e^+\nu)}{
 2\pi^2\sin^4\theta_W}\lambda^8=(5.26\pm 0.06)\cdot 10^{-11}
\left[\frac{\lambda}{0.225}\right]^8.
\end{equation}
 and 
\begin{equation}\label{bklpn}
\mbox{BR}(K_L) \equiv \mbox{BR}(K_L\to\pi^0\nu\bar\nu)=\kappa_L\cdot
\left({\rm Im} \left(\frac{\lambda_t}{\lambda^5}X(x_t)\right)\right)^2
\end{equation}
\begin{equation}\label{kapl2}
\kappa_L=\kappa_+ \frac{r_{K_L}}{r_{K^+}}
\frac{\tau(K_L)}{\tau(K^+)}=
(2.29\pm 0.03)\cdot 10^{-10}\left[\frac{\lambda}{0.225}\right]^8
\end{equation}
The numbers for $r_{K_L}$ and $r_{K^+}$, describing the isospin breaking effects to the $K_{l3}$ decay, have recently been updated in \cite{Isidori:2006qy}.
Due to the absence of the charm contribution, $\klpn$ is theoretically even cleaner than $\kpn$.
Turning now to the contributions from new physics, we find that in both cases the leading term stems from a tree diagram transmitted by the $Z'$ boson.
For the effective Hamiltonian, this leads to a new term of the form
\begin{equation}
H_{eff}^{Z'}= \sum_{l=e,\mu,\tau} \frac{G_F}{\sqrt{2}} \frac{\tilde V_{32}^* \tilde V_{31}}{3} \left( \frac{M_Z c_W}{M_{Z'}} \right)^2 (\bar s d)_{V-A}
(\bar \nu_l  \nu_l)_{V-A}\,.
\end{equation} 
This can be absorbed into the modification of the function $X(x_t)$ as
\begin{equation}
\Delta X_{K \pi \nu \nu}=\frac{s_W^2 c_W^2}{\alpha} \frac{2 \pi}{3} \frac{\tilde V_{32}^* \tilde V_{31} }{V_{ts}^* V_{td}} \left( \frac{M_Z}{M_Z'} \right)^2\,.
\end{equation}
We have already written (\ref{bkpnn}) and (\ref{bklpn}) in such a way that using the thus modified function $X(x_t)$ gives the correct branching ratio in 
the 331 model. 

The present experimental situation of these decays can be summarized as follows \cite{Anisimovsky:2004hr,Ahn:2006uf}:
\begin{equation}
\mbox{BR}(\kpn)=(14.7^{+13.0}_{-8.9})\cdot 10^{^{-11}}, \qquad \mbox{BR}(\klpn)<2.1 \cdot 10^{-7} \quad (90\% \mbox{CL}) \,,
\end{equation} 
while the SM can be quoted as \cite{Buras:2006gb}
\begin{equation}
\mbox{BR}(\kpn)=(8.0 \pm 1.1)\cdot 10^{^{-11}}, \qquad \mbox{BR}(\klpn)=(2.9 \pm 0.4) \cdot 10^{-11}  \,.
\end{equation} 

Turning next to $B_{d/s} \to \mu^+ \mu^-$, the SM effective Hamiltonian is given by
\begin{equation}
H_{eff}^{B_{d/s}\mu \mu}= -\frac{G_F}{\sqrt{2}}\frac{\alpha}{2 \pi s_W^2} (V^*_{tb}V_{td/s}) Y(x_t) (\bar b q)_{V-A}(\bar \mu \mu)_{V-A}\,,
\end{equation}
which leads to the following formulae for the branching fractions:
\begin{equation}\label{bblls}
\mbox{BR}(B_q\to \mu^+\mu^-)=
\tau_{B_q} \frac{G_F^2}{\pi} m_{B_q}
\left(\frac{\alpha F_{B_q} m_{\mu}}{4 \pi \sin^2 \theta_W} \right)^2 \sqrt{1-4 \frac{m_{\mu}^2}{m_{B_q}^2}}
|V^\ast_{tb}V_{tq} Y(x_t)|^2
\end{equation}
%\begin{equation}\label{bblld}
%\mbox{BR}(B_d\to \mu^+\mu^-)=1.82\times 10^{-6} \times
%\left[\frac{\tau_{B_d}}{1.54~{\rm ps}}\right]
%\left[\frac{F_{B_d}}{203~ {\rm MeV}}\right]^2
%|V^\ast_{tb}V_{td} Y(x_t)|^2,
%\end{equation}
Due to the uncertainties in the decay constants, these decays are theoretically slightly less clean than the $K \to \pi \nu \bar \nu$ decays. 
Similarly to the $K \to \pi \nu \bar \nu$ decays, the new contribution to $B_{d/s} \to \mu^+ \mu^-$ is given by:
 \begin{eqnarray}
H_{eff}^{Z'}&=& \frac{G_F}{\sqrt{2}} \frac{\tilde V_{33}^* \tilde V_{31/32}}{3} \left( \frac{M_Z c_W}{M_{Z'}} \right)^2 (\bar b q)_{V-A}(\bar \mu \mu)_{V-A}+ \\
            & & \frac{G_F}{\sqrt{2}} \frac{2 \tilde V_{33}^* \tilde V_{31/32}}{3} \left( \frac{M_Z c_W}{M_{Z'}} \right)^2 (\bar b q)_{V-A}(\bar \mu \mu)_{V+A}
\end{eqnarray}
Since only the axial-vector component in the lepton current contributes to the decay, we can project the $V+A$ contribution onto the $V-A$ one to arrive at
\begin{equation}
H_{eff}^{Z'}= -\frac{G_F}{\sqrt{2}} \frac{\tilde V_{33}^* \tilde V_{31/32}}{3} \left( \frac{M_Z c_W}{M_{Z'}} \right)^2 (\bar b q)_{V-A}(\bar \mu \mu)_{V-A}
\end{equation}
and the modification in $Y(x_t)$ is 
\begin{equation}
\Delta Y_{B\mu \mu}=\frac{s_W^2 c_W^2}{\alpha} \frac{2 \pi}{3} \frac{\tilde V_{33}^* \tilde V_{31/32} }{V_{tb}^* V_{td/ts}} \left( \frac{M_Z}{M_Z'} 
\right)^2 \,.
\end{equation}
Again, (\ref{bblls}) is written in such a way that the modification of $Y(x_t)$ leads to the correct result in the 331 model. 
The experimental bounds on these decays read as \cite{Bmumu}
\begin{equation}
\mbox{BR}(B_s\to \mu^+\mu^-)< 1\cdot 10^{-7}\qquad \mbox{BR}(B_d\to \mu^+\mu^-)<3 \cdot 10^{-8}  \quad (90\% \mbox{CL}) \,,
\end{equation} 
where the most recent SM predictions are \cite{Blanke:2006ig}
\begin{equation}
\mbox{BR}(B_s\to \mu^+\mu^-)=(3.35\pm0.32)\cdot 10^{-9}   \qquad \mbox{BR}(B_d\to \mu^+\mu^-)=(1.03 \pm 0.09) \cdot 10^{-10} \,.
\end{equation}

Finally, we give also the contributions to the decay $K_L \to \pi^0 e^+ e^-$. 
In the SM, the short-distance CP violating part of the effective Hamiltonian is given at tree-level (of the matrix elements) by:
\begin{equation}
H_{eff}^{K\pi ll}=-\frac{G_F}{\sqrt{2}} V_{ts}^* V_{td}(y_{7V} Q_{7V}+y_{7A} Q_{7A}) \,, 
\end{equation}
where  $Q_{7V}=(\bar s d)_{V-A} \bar e\gamma^{\mu}e$ and $Q_{7A}=(\bar sd)_{V-A}\bar e\gamma^{\mu} \gamma^5e$ are the vector- and axial-vector operators contributing, while
 the matching conditions of the Wilson coefficients $y_{7V}$ and $y_{7A}$ are
\begin{equation}
y_{7V}=\frac{\alpha}{2 \pi} \left( \frac{Y_0(x_t)}{s_W^2} -4 Z_0(x_t)+P_0\right) 
\end{equation}
\begin{equation}
y_{7A}=-\frac{\alpha}{2 \pi} \frac{Y_0(x_t)}{s_W^2}
\end{equation}
Here we have followed the normalizations in \cite{Buchalla:2003sj,Isidori:2004rb}, $P_0=2.89\pm0.06$ and have neglected a small term $P_E.$ 

In principle, the NP amplitude here is given just as in the case of $K \to \pi \nu \bar \nu$ 
by a tree-level $Z'$ exchange, but, in this case there is also a right-handed contribution, leading the complete amplitude to be
\begin{equation}
H_{eff}^{Z'}=\frac{G_F}{\sqrt{2}} \left( \frac{M_Z c_W}{M_Z'} \right)^2 \left(Q_{7V}+\frac{1}{3}Q_{7A} \right) (\tilde V_{32}^* \tilde V_{31})\,.
\end{equation}
Instead of absorbing these new contributions into modifications of the Inami Lim Functions, we will here absorb them into the matching conditions of the Wilson 
coefficients\footnote{This is because, due to the existence of right- and left-handed couplings, it is not possible to define one {\em universal} $C$ function 
for this decay.}, i.e.
\begin{eqnarray}
\Delta y_A &=& -\frac{1}{3} \left( \frac{M_Z c_W}{M_Z'} \right)^2  \frac{(\tilde V_{32}^* \tilde V_{31})}{V_{ts}^* V_{td}}\\
\Delta y_V &=& - \left( \frac{M_Z c_W}{M_Z'} \right)^2 \frac{(\tilde V_{32}^* \tilde V_{31})}{V_{ts}^* V_{td}}
\end{eqnarray}
We refrain from giving the complete formula for the branching ratios since these are rather lengthy, and refer the reader to \cite{Mescia:2006jd}, for the explicit 
expressions, including also the long-distance indirectly CP-violating terms and their interference with the short-distance contributions.  
Finally, let us also here quote the corresponding SM predictions and current experimental limits. They are \cite{Mescia:2006jd}
\begin{equation}
\mbox{BR}(K_L \to \pi^0 e^+ e^-)=(3.54^{+0.98}_{-0.85}) \cdot 10^{-11} \,, \qquad \mbox{BR}(K_L \to \pi^0 \mu^+ \mu^-)= (1.41^{+0.28}_{-0.26}) \cdot 10^{-11} \,,
\end{equation}
and  \cite{Alavi-Harati:2003mr,Alavi-Harati:2000hs}
\begin{equation}
\mbox{BR}(K_L \to \pi^0 e^+ e^-)< 28 \cdot 10^{-11} \,, \qquad \mbox{BR}(K_L \to \pi^0 \mu^+ \mu^-)< 38 \cdot 10^{-11}  \quad (90\% \mbox{CL}) \,,
\end{equation}
respectively, where the SM prediction corresponds to positive interference between mixed and direct CP violation, which is favored \cite{Buchalla:2003sj,Friot:2004yr}. 

\section{Numerical Analysis}\label{sec:numerics}
\subsection{General Remarks}
In this section, we analyze numerically the expressions given in the previous section. 
Before we do so, let us briefly review the framework and give the input we use.
The CKM matrix is constructed from the measurements of tree level dominated decays, namely the experimental values of the Unitarity Triangle (UT) side $R_b$ as 
determined from the measured values of $|V_{ub}|$ and $|V_{cb}|$, as well as $|V_{us}|$ and the UT angle$\gamma$. As we have seen above, all further  
constraints may be polluted by new contributions from $Z'$ 
exchange\footnote{Tree level contributions analogous to the appearing in muon decay are not possible, since we are
dealing with charged quark transitions here.}. 
To be specific, the tree level extraction for $\gamma$ from $B \to D^{(*)} K$ decays leads to 
\begin{equation}
\gamma=(71 \pm 16)^{\circ}, \qquad \gamma=-(109\pm 16)^{\circ} \,,
\end{equation}
where there is a two-fold ambiguity in this determination of  $\gamma$, with the second solution in 
contradiction to the SM. This solution is disfavored from the combination of $\cos (2 \beta +\phi_d)$ and the semileptonic asymmetries $A^{d/s}_{SL}$
\cite{Bona:2006sa}. Therefore, we will work only with the first solution and construct our unitarity triangle from it.
The further input values are collected in 
Table \ref{tab:input} \cite{Blanke:2006eb}. The values of $|V_{ub}|$ and $|V_{cb}|$ are obtained from an average of both inclusive and exclusive determinations. Note, that 
obtaining the "SM-predictions" in the following by setting the $X(x_t)$ and $Y(x_t)$ functions to their SM values leads to different predictions for the
decay rates than the SM predictions quoted above. This is due to the different CKM factors used, since we are working with only the tree-level input 
parameters, while the earlier SM predictions used all input available in the UT fit.

\begin{table}[ht]
\renewcommand{\arraystretch}{1}\setlength{\arraycolsep}{1pt}
\center{\begin{tabular}{|l|l|}
\hline
{\small $G_F^{\mu}=1.16637\cdot 10^{-5} \gev^{-2}$} & {\small$\Delta M_K= 3.483(6)\cdot 10^{-15}\gev$} \\
{\small$M_W= 80.425(38)\gev$} & {\small$\Delta M_d=0.507(4)/ \rm{ps}$\hfill} \\\cline{2-2}
{\small$\alpha=1/127.9$} &{\small $\Delta M_s = 17.77(12)/ \rm{ps}$\hfill} \\\cline{2-2}
{\small$\sin^2 \theta_W=0.23120(15)$\qquad\hfill} & \\\hline
{\small$|V_{ub}|=0.00409(25)$} & {\small $F_K\sqrt{\hat B_K}= 143(7)\mev$\qquad\hfill} \\\cline{2-2}
{\small $V_{cb} = 0.0416(7)$\hfill} & {\small$F_{B_d} \sqrt{\hat B_{B_d}}= 213(38)\mev$}\\\cline{1-1}
{\small$\lambda=|V_{us}|=0.225(1)$ \hfill} & {\small$F_{B_s} \sqrt{\hat B_{B_s}}= 262(35)\mev$} \\\hline
 {\small$|V_{ts}|=0.0409(9)$ \hfill} & {\small$\eta_1=1.32(32)$\hfill} \\\cline{1-1}
{\small$m_{K^0}= 497.65(2)\mev$} & {\small$\eta_3=0.47(5)$\hfill}\\%\cline{2-2}
%{\small$m_{D^0}=  1.8645(4)\gev$} &{\small$\eta_2=0.57(1)$} \\
{\small$m_{B_d}= 5.2794(5)\gev$} & {\small$\eta_2=0.57(1)$}\\\cline{2-2}
{\small$m_{B_s}= 5.375(2)\gev$} &{\small$\eta_B=0.55(1)$\hfill} \\\cline{2-2}
{\small $|\varepsilon_K|=2.284(14)\cdot 10^{-3}$ \hfill} &  {\small$ \bar m_c= 1.30(5)\gev$}\\
\cline{1-1}
{\small $\beta=25.1(2.1)^{\circ}$ \hfill} & {\small$ \bar m_t= 163.8(32)\gev$}\\
\hline
\end{tabular}  }
\caption {Values of the experimental and theoretical
    quantities used as input parameters. The value of $\beta$ is the number we find when constructing the UT from the other input \cite{Yao:2006px, HFAG}.}
\label{tab:input}
\renewcommand{\arraystretch}{1.0}
\end{table}

We perform the subsequent numerical analysis in two steps:
\begin{itemize}
\item In the first step, we consider the observables $\Delta M_K$, $\Delta M_{d/s}$, $\varepsilon_K$ and $\sin 2\beta$. All 
these quantities are related in some way to $K^0 - \bar K^0$ and $B^0_q - \bar B^0_q$ mixing, and have been measured to a significant 
precision. Therefore, we can use them to constrain the parameter space of the minimal 331 model. In this context, we also study the
$B_s^0 - \bar B_s^0$ mixing phase $\beta_s$, that can, in principle, be measured through $A^{\rm mix}_{CP}(B^0_s \to J/\psi \phi)$, but is, as of
yet, unknown.
\item In the second step, we study the implications of these bounds for several rare decays, in particular the decays 
$\kpn$, $\klpn$ and $B_{d/s} \to \mu^+\mu^-$. In this context, we are mainly interested in obtaining potential upper
bounds for these decays, as well as in finding correlations that would allow an unambiguous test of the model
\end{itemize}

\subsection{Constraints from $\Delta M_K$, $\varepsilon_K$ and $B_q^0 - \bar B_q^0$ Mixing}
In this subsection, we focus on the bounds on the model that can already be obtained by studying well-measured quantities.
If one considers the theoretical expressions, there are always several parameters appearing in the corresponding bounds,
i.e. the mass of the $Z'$ boson, as well as the corresponding combination of mixing-matrix elements. Therefore, one can 
now pursue two possible analyses: The first possibility, which has been followed repeatedly in the literature 
\cite{GomezDumm:1994tz,Rodriguez:2004mw}, is to assume
a certain texture of the mixing matrix (in most cases this has been assumed to have a Fritzsch-type structure, while another texture has been used in 
\cite{Diaz:2004fs}), which
then allows to obtain bounds on the $Z'$ mass. Several times, this has led to bounds that are potentially conflicting with the 
upper bounds obtained from the Landau Pole. On the other hand \cite{Liu:1993gy}, one can set $M_{Z'}$ onto this upper bound 
and thereby obtain some information on the size of the corresponding mixing matrix elements. In order to be able to deal with the most general 
situation, we prefer not to make use of any specific parameterization of the mixing matrix, but rather follow the second possible approach, in a somewhat
more general manner when considering the implications for rare decays. For the moment, we fix the $Z'$ mass to $M_{Z'}=1 ~\mathrm{TeV}$ and 
$M_{Z'}=5 ~\mathrm{TeV}$ as two representative values which give us bounds on the real and imaginary parts of $(V_{31} V_{32}^*)^2$ if we 
consider the bounds from $\Delta M_K$ and $\varepsilon_K$, respectively. On the technical level, we proceed in a manner
that is inspired by the analysis \cite{Blanke:2006eb} of the Littlest Higgs model with T-parity, where the uncertainties in the theoretical input are absorbed 
into a generously assigned experimental error. We use, as possible deviations from the central value, $40\%$ for $\Delta M_{d/s}$ as well as for $\varepsilon_K$, 
$50 \%$ for $\Delta M_K$ 
and $4^{\circ}$ for $\beta$. These $4^{\circ}$ correspond to an uncertainty of about $8\%$ in $\sin 2 \beta$, as in 
\cite{Blanke:2006eb}\footnote{While we use a somewhat newer experimental value of $\Delta M_s$ (the number from \cite{Blanke:2006eb} is $\Delta M_s = 17.7 \pm 0.4$), 
we choose to retain the assigned percentage of the uncertainty due to the fact that the theoretical error vastly dominates (0.4 are only $2\%$ of 17.4).}.
A slight modification of $\sin 2 \beta$ would certainly be welcome in view of small discrepancy between the value of $\sin 2 \beta$ from $B \to  J/\psi K_S$ and the one
obtained from a UT fit without this input. This discrepancy can be attributed to a small experimental number $\sin 2 \beta |_{J/\psi K_S}$ or a
large value of $|V_{ub}/V_{cb}|$. We also keep the CKM parameters fixed at their central values since we are mainly interested in the effects that are induced by
new physics, not those that arise from parameter variation.

\begin{figure}
\begin{center}
\includegraphics{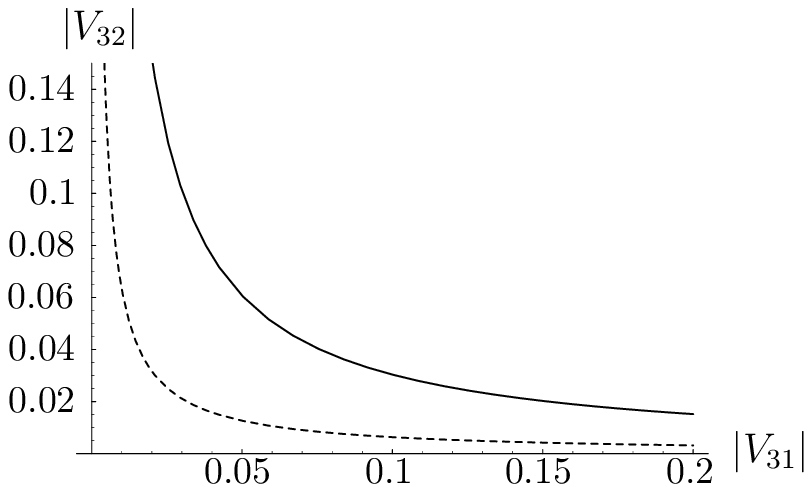}
\end{center}
\caption{\label{V32MK} An upper bound on $|V_{32}|$ coming from $\Delta M_K$ for $M_{Z'}=1~\mathrm{TeV}$(dotted) and $M_{Z'}=5~\mathrm{TeV}$(solid).}
\end{figure}

We find (taking $M_{Z'}=5 ~\mathrm{TeV}$ for definiteness - a similar pattern shows for other values) $\mathrm{Re}[(V_{31} V_{32}^*)^2]<9.2 \cdot 10^{-6}$ and 
$\mathrm{Im}[(V_{31} V_{32}^*)^2]<4.8 \cdot 10^{-8}$, from which we can conclude that the imaginary part of this amplitude is much stronger constrained than the real part. 
Therefore, if we would like to saturate the bounds, we should consider an entirely real or entirely imaginary value  of $V_{31} V_{32}^*$.  
There is then, from $\Delta M_K$, a bound on  $|V_{32}|$ which depends on $|V_{31}|$, as shown in Fig. \ref{V32MK}. Notice, that both elements $|V_{31}|$ and 
$|V_{32}|$ can not be simultaneously large. This is true for both of the chosen values of $M_{Z'}$ that we are showing.
%\begin{figure}
%\begin{center}
%\includegraphics{Scen0/V33_Sin2b.eps}
%\end{center}
%\caption{\label{V33S2b}The upper bound on $V_{33}$ coming from $\sin 2 \beta$}
%\end{figure}

\begin{figure}
\begin{center}
\includegraphics{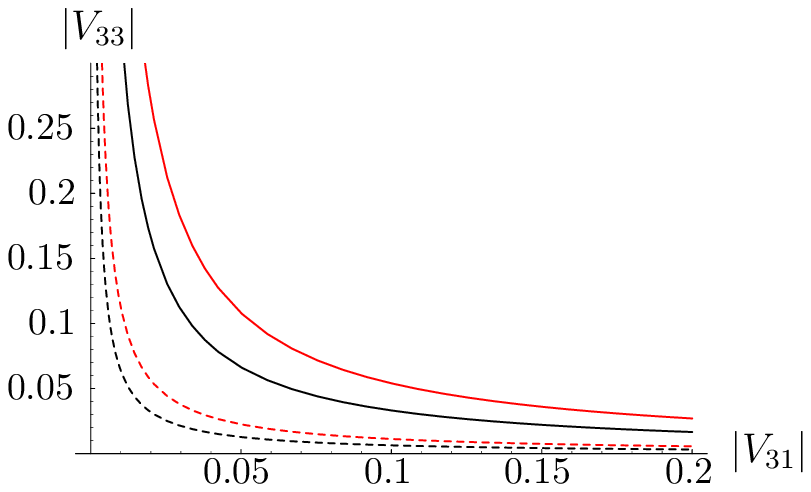}
\end{center}
\caption{\label{V33dMBd}The upper bound on $|V_{33}|$ coming from $\Delta M_d$ assuming that the SM and NP contributions are aligned (red), and the complementary bound 
from $\sin 2 \beta$ if they are perpendicular (black). Both are given for $M_{Z'}=1~\mathrm{TeV}$ (dotted) and $M_{Z'}=5~\mathrm{TeV}$(solid).}
\end{figure}

The corresponding bounds from $B^0_q - \bar B^0_q$ mixing are somewhat more subtle to deal with, since the new contributions are not simply added here, so that also 
interference terms are important. An estimate of the bounds can be obtained by assuming
\begin{itemize}
\item That the new contributions and the SM one are directly aligned, where then the deviation from the SM corresponds directly to the new contributions, or
\item That the new contribution is constructed in such a way that it is perpendicular to the SM contributions in the complex plain, i.e. comes with a phase 
$\beta/\beta_s \pm 90^{\circ}.$
\end{itemize}
In the first case, the absolute value of the new contribution is minimal, while in the second it is maximal. On the other hand, taking an aligned contribution
allows one to circumvent the bounds coming from $\sin 2 \beta |_{J/\psi K_S}$, which is much more stringent than the one from $\Delta M_d$. 
To show the complementarity of the two bounds, we plot, in Fig.\ref{V33dMBd}, the bound coming from $\Delta M_d$ in the case of aligned contributions and 
the bound coming from $\sin 2\beta$ in the case of orthogonal contributions. We find that the bound from $\sin 2\beta$
is stronger than the one from $\Delta M_d$ so that a contribution that is aligned with the SM one can be larger.

On the other hand, the mixing phase $\beta_s$ has not been measured, and can not be used to constrain the combination $V_{32} V_{33}^*$. This has two 
implications:
\begin{itemize}
\item First, we can have, in this case, a "mirror solution" of $\Delta M_s$ in which the new contribution is antiparallel to the SM, but twice as large.
      At present, this situation can not be excluded with the observables we are studying. It is, however, possible, that the large absolute value of the
      this new new contribution would violate bounds from $b \to s~ \gamma$. The new contributions to 
      $b \to s~ \gamma$ are, however, loop suppressed, not only through the arising couplings, but additionally by heavy propagators, so that we expect 
      the influence to be only marginal.
\item There is no strong bound on an "orthogonal contribution", which means that the phase $\beta_s$, as measured through 
      $A_{CP}^{\rm mix}(B_s^0 \to J/\psi \phi)$, may be rather large. In fact, we find that the present range of $\beta_s$ is entirely unconstrained, since there is an 
      allowed range connecting the mirror solution with the SM-like ones. Clearly, a measurement of this phase would severely constrain the available parameter space. 
\end{itemize}

%Therefore, the largest 
%absolute value of this combination is allowed, when both contributions are perpendicular to each other, leading to a modification of the mixing phase. Taking
%the combination $|V_{32} V_{33}^*|/M_{Z'}$ at the upper bound implied by $\Delta M_s$, we find that a value of $\Delta \beta_s=22^{\circ}$ would be reached. At the
%level of the CP asymmetry of $B_s^0 \to J/\psi \phi$, which measures $\sin 2 \beta_s$ in the SM, this corresponds to numbers as large as 
%$A_{CP}^{\rm mix}(B_s^0 \to J/\psi \phi)=0.75$. This bound depends only very slightly on $M_{Z'}$ through the corresponding dependence of $G_F$.

Finally, we point out that both $\Delta M_s$ and $\Delta M_d$ can be equally well enhanced or suppressed, since the sign of the new contributions can be simply
switched by a change of sign in the mixing matrix $\tilde V_L$, so that no preferred behavior of the prediction can be obtained. On the other hand, if the data in 
either process should indicate an enhancement or suppression, it could always be satisfied within the minimal 331 model.

\subsection{Implications for Rare Decays}
Let us now study the implications of the bounds derived above for the modification in rare decay amplitudes. The strategy of the analysis will be to 
saturate the bounds by fixing the corresponding combination $(V_{ij} V_{kl})/M^2_{Z'}$, thereby leaving $M_{Z'}$ as the only variable left in the
expressions for the rare decays. In this way, we find upper bounds for the rare decays as a function of the mass of the $Z'$ boson. For an earlier study of 
$\kpn$ in the 331 model, see \cite{Long:2001bc}. Our analysis goes beyond that one in that we consider not only the tree-level process but also the 
one-loop SM amplitude and the interference between the two. This is definitely appropriate, since the SM is expected to be the main contribution in most FCNC
processes.

Beginning with the rare $K$ decays, we can use the information obtained from the previous section, that the real part of $((V_{ij} V_{kl})/M_{Z'})^2$ is much 
less constrained than the imaginary part, so that we set:
\begin{equation}
\mathrm{Re}[(V_{ij} V_{kl})^2]=(\mathrm{Re}[V_{ij} V_{kl}])^2\,,
\end{equation} 
which effectively amounts to setting the imaginary part to zero. Alternatively, one could set $\mathrm{Re}[V_{ij} V_{kl}]=0$, where then the new contribution
is purely imaginary. We will discuss this setup when we look at $\klpn$ in more detail. For $\kpn$, however, we will indeed be concerned only with the purely 
CP conserving case.

\begin{figure}
\begin{center}
\includegraphics[height=6cm]{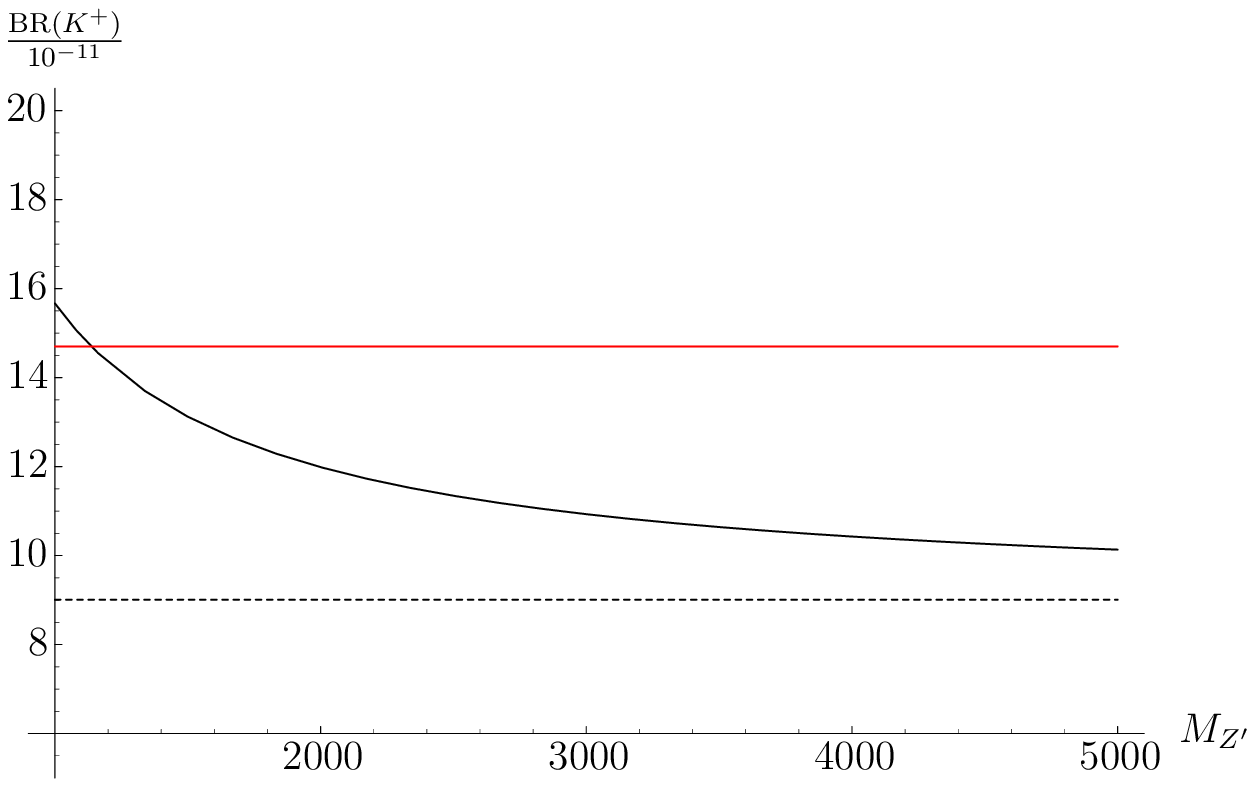}
\end{center}
\caption{\label{KplMZpr} Upper bound on the decay $K^+ \to \pi^+ \bar \nu \nu$ taking into account the constraints from $\Delta M_K$. The SM value is 
denoted with a dashed line, while the present experimental central value is given by the red line.}
\end{figure}

Proceeding in this manner, we find an upper bound on $K^+ \to \pi^+ \bar \nu \nu$ as shown in Fig. \ref{KplMZpr}. In addition, we give also the central 
value of the experimental result \mbox{$\mathrm{BR}(K^+)=14.7^{+13}_{-8.9} \cdot 10^{-11}$}.  This experimental measurement is above the SM prediction, but well 
compatible within theoretical and experimental uncertainties. We find, that only rather low values of the $Z'$ mass can reach this central number. 

\begin{figure}
\begin{center}
\includegraphics[height=6cm]{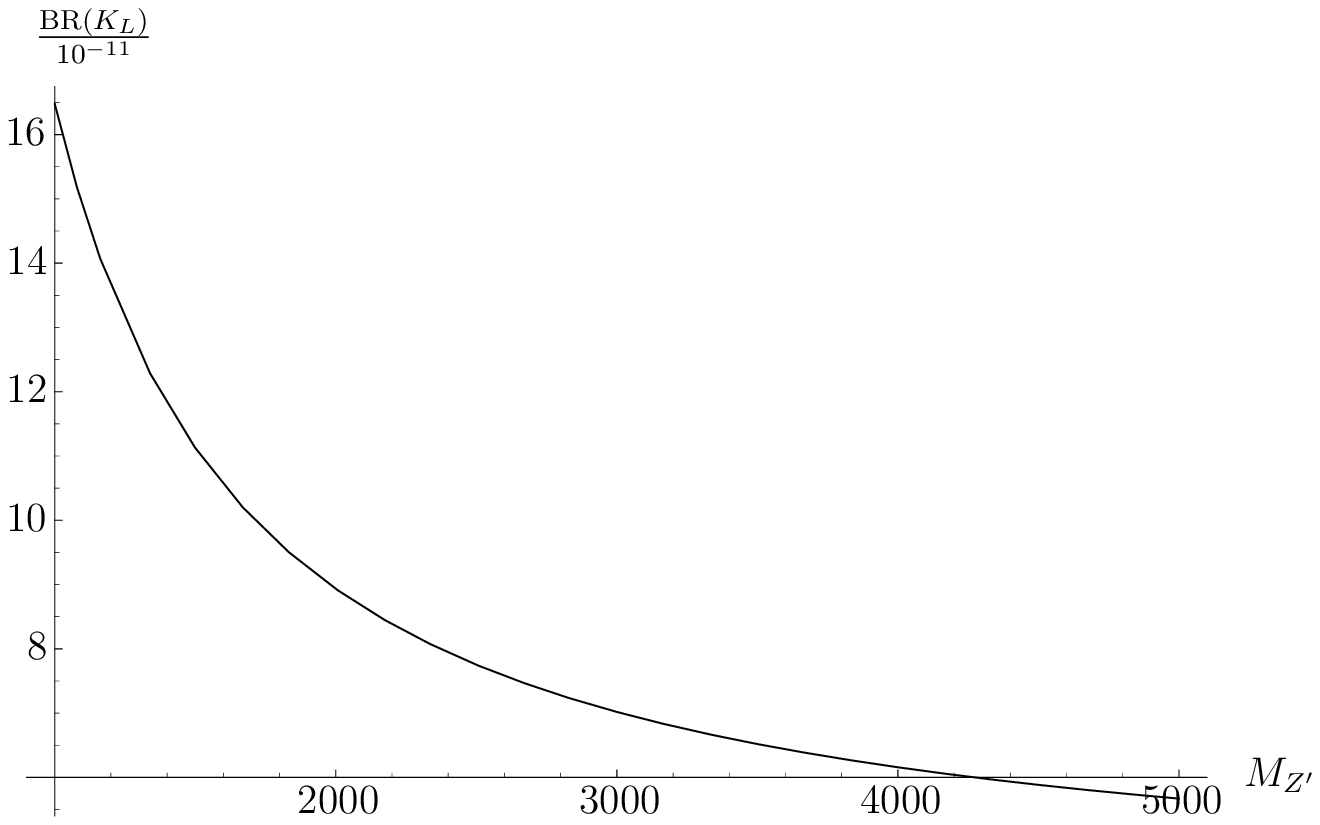}
\end{center}
\caption{\label{KLMZpr} Upper bound on the decay $K_L \to \pi^0 \bar \nu \nu$ taking into account the constraints from $\Delta M_K$ in the case of a purely CP
violating $Z'$ contribution. In case of CP conserving $Z'$ couplings, this bound becomes more stringent.}
\end{figure}

Concerning the decay $K_L \to \pi^0 \nu \bar \nu$, we find it most instructive to show the upper bound that is obtained in the case of a purely CP violating $Z'$
contribution. In this case, the bound for $\mathrm{Im}(V_{31} V_{32})$ is also given by the bound in $\Delta M_K$, and is shown in Fig. \ref{KLMZpr}. Again, we find that 
large enhancements are, in  principle, possible, in particular for values of the $Z'$ mass that lie beneath about $2 \mathrm{TeV}$ . Therefore, it is clear that 
visible signals in both $K \to \pi \bar \nu \nu $ decays can still be expected. In particular, values such as the current experimental central value of $\kpn$ are 
entirely possible.

We have, in both cases, not shown the possibility of a suppression of both branching fractions. 

\begin{figure}
\begin{center}
\includegraphics[height=7cm]{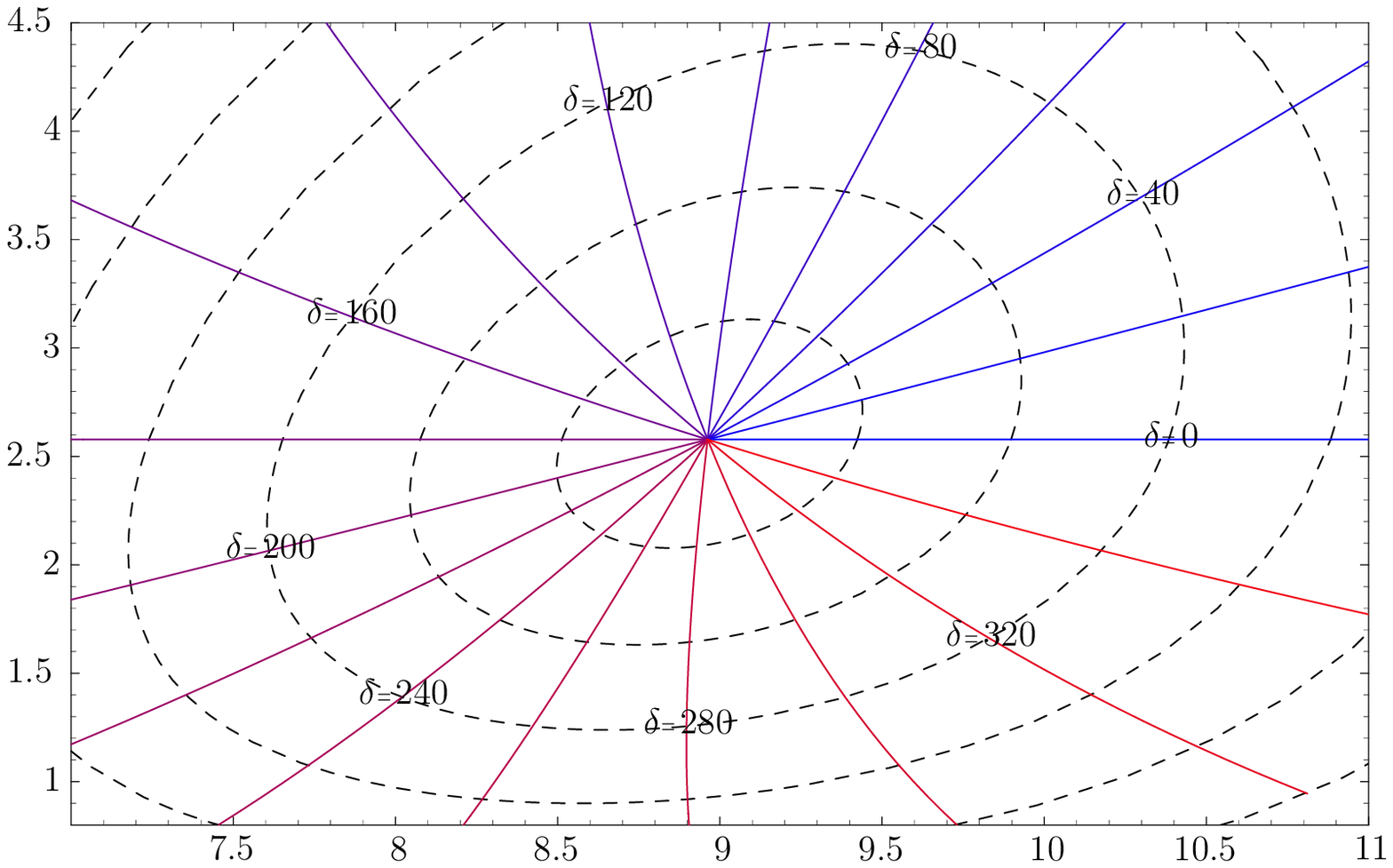}
\end{center}
\caption{\label{KLKP}A projection onto the $\klpn$-$\kpn$ plane. Measuring both branching fraction allows to unambiguously determine both the phase as well
as the magnitude of the new physics contribution. We vary $A\equiv (\tilde V_{31} \tilde V_{32})/M_{Z'}^2$ as \mbox{$A=(0-30) \cdot 10^{-11}$} in steps of $5 \cdot 10^{-11}$.}
\end{figure}

In addition, as discussed in a slightly different context in \cite{Buras:2004ub}, a measurement of both decays is sufficient to find both the absolute value and 
phase of the unknown quantity $A\equiv (\tilde V_{31} \tilde V_{32})/M_{Z'}^2$, along the
lines of Fig \ref{KLKP}. Here, the dashed circles correspond to variations of the phase for various values of $A$, while the colored rays correspond to 
fixed values of the phase $\delta_{12} \equiv \delta_2-\delta_1$. We show here only a restricted area of the possible branching fractions, 
but it is clear that a measurement of both decays uniquely fixes all parameters in question.

\begin{figure}
\begin{center}
\includegraphics[height=7cm]{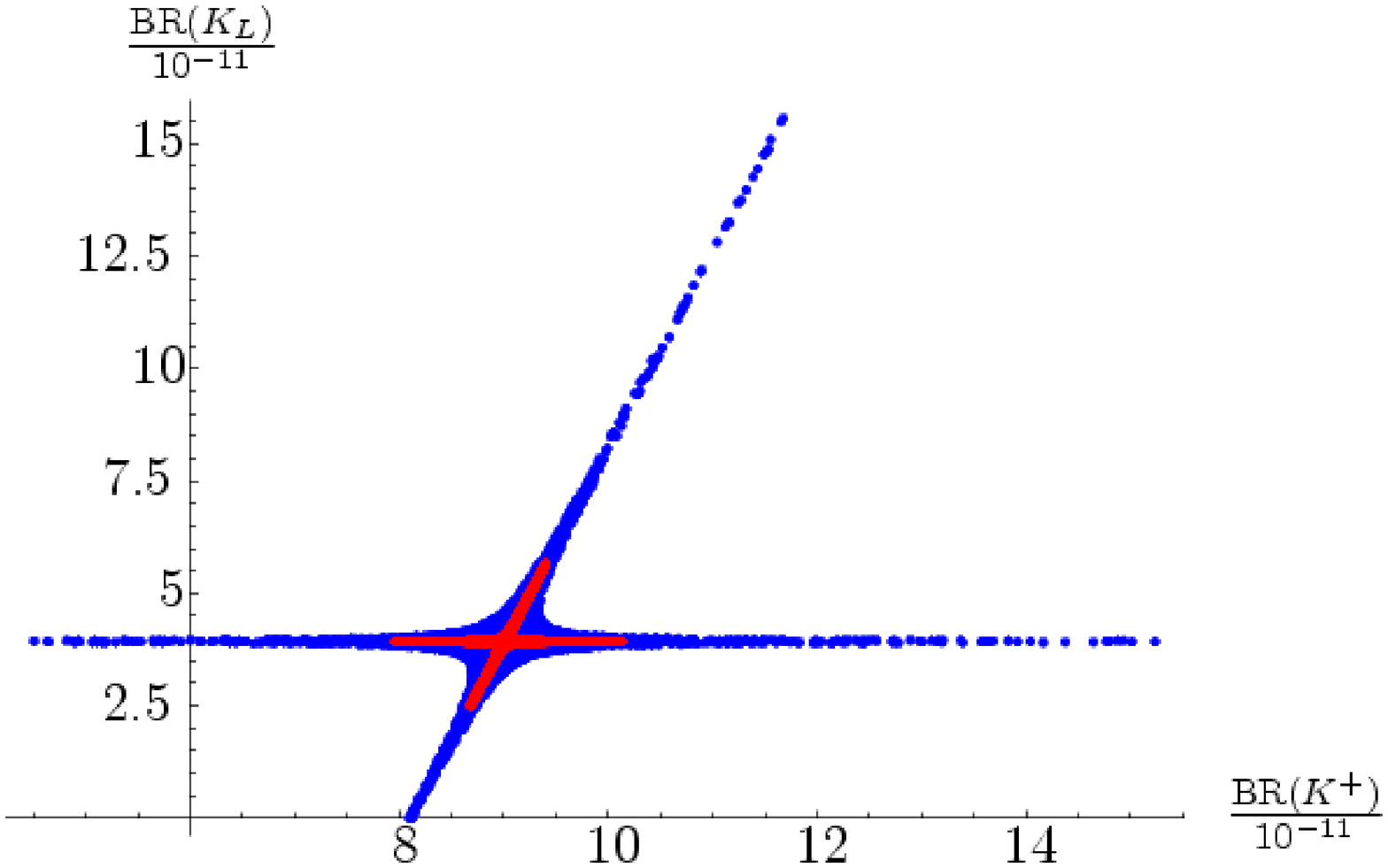}
\end{center}
\caption{\label{KLKPscat}A projection onto the $\klpn$-$\kpn$ plane including the upper bounds from $\Delta M_K$ and $\epsilon_K$ for $M_{Z'}=5~\mathrm{TeV}$ (red)
and  $M_{Z'}=1~\mathrm{TeV}$ (blue).}
\end{figure}

In this context, it is interesting to see, which values of the branching fractions are actually allowed through the bounds coming from $\Delta M_K$ and $\varepsilon_K$.
Therefore, we now show again the $\klpn$-$\kpn$ plane in Fig. \ref{KLKPscat} with those areas cut out, which are ruled out by the respective bounds. Here, the red 
star corresponds to $M_Z'=5~\mathrm{TeV}$, while the blue star shows those values that are allowed for $M_Z'=1~\mathrm{TeV}$. Notice that there are, similar to
the pattern seen in the Littlest Higgs model with T-parity \cite{Blanke:2006eb}, several branches in this plane that are allowed. This is due exactly to the effect 
mentioned above, namely that the bound on $\varepsilon_K$ is stronger than the one on $\Delta M_K$, and the branches correspond to those areas, where the phase of the 
new contributions is such that it doesn't modify $\varepsilon_K$ strongly. This is nicely seen in the comparison of both Figures in the $\kpn-\klpn$ plane, and should
actually be a general effect of any model. Notice, that, due to the leptophobic character of the $Z'$ boson, the possible modifications of both branching fractions are 
not very large in comparison to the LHT model. Also, this figure nicely demonstrates how the allowed region decreases as the $Z'$ mass is increased.

\begin{figure}
\begin{center}
\includegraphics[height=4.7cm]{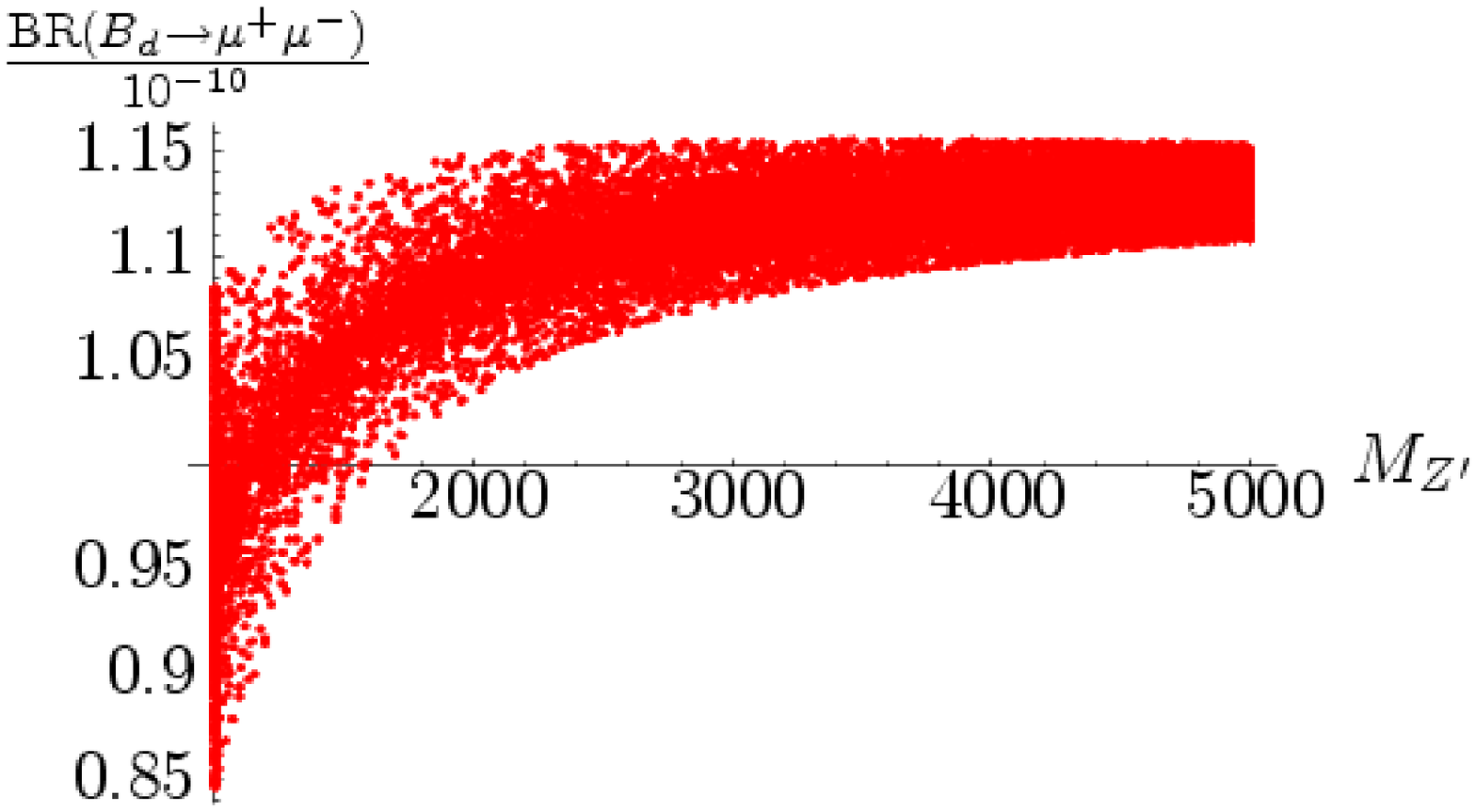}
\includegraphics[height=4.7cm]{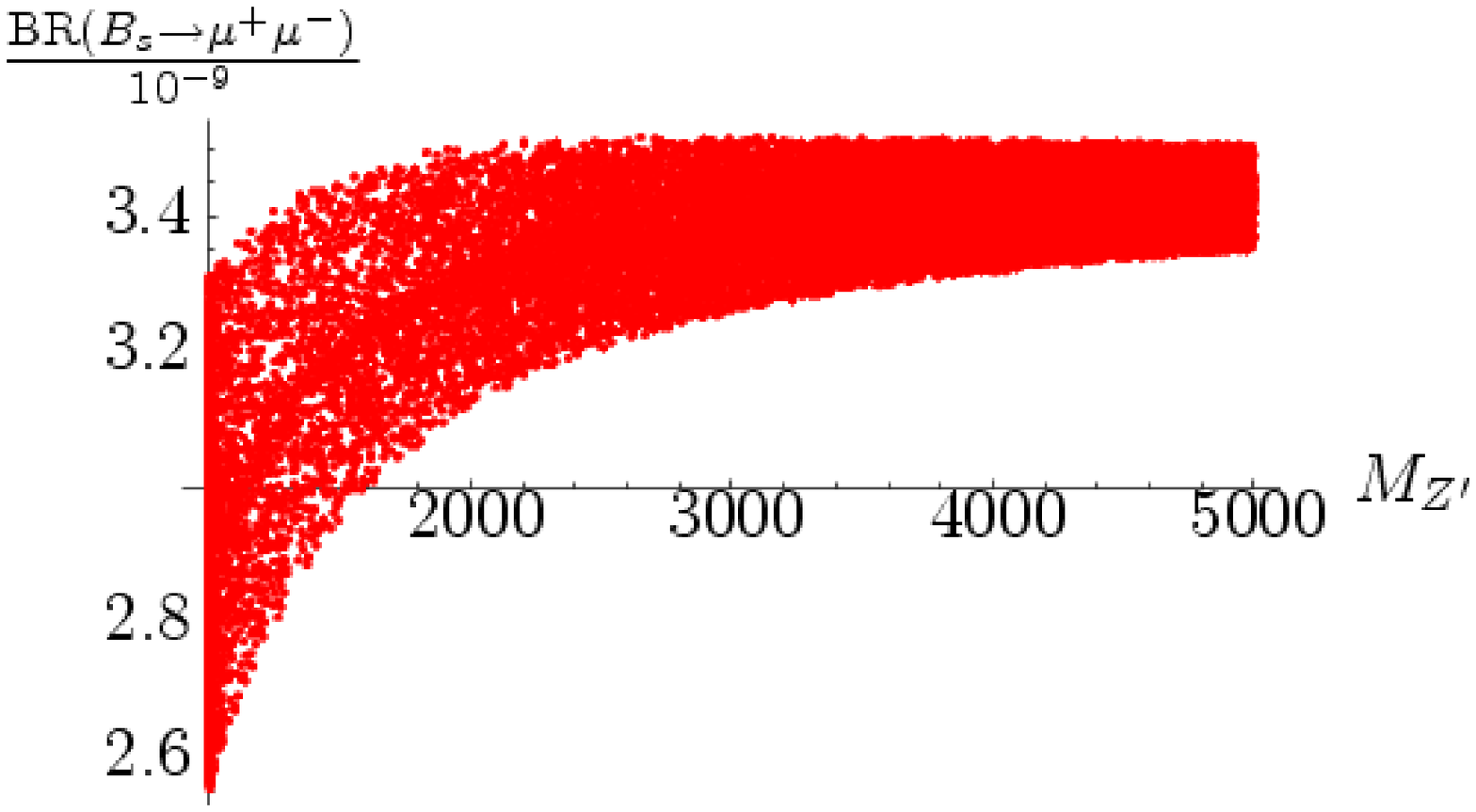}
\end{center}
\caption{\label{bdmumu}Allowed range for the branching ratio $B_d \to \mu^+ \mu^-$ obtained from $\Delta M_d$ and $\sin 2 \beta$ and for the branching ratio 
$B_s \to \mu^+ \mu^-$ from $\Delta M_s$ .}
\end{figure}

Let us now turn to the decays $B_{d/s}\to \mu^+ \mu^-$. Here, we can use the bounds coming from $\Delta M_d$ and from $\sin 2 \beta$ to obtain an upper bound on the
branching ratio $B_{d}\to \mu^+ \mu^-$. 
The result of this exercise is shown in Fig. \ref{bdmumu}. Interestingly, this result makes a suppression of the branching ratio seem much more 
likely than an enhancement, and, in any case, a strong enhancement of this branching ratio would unambiguously rule out the minimal 331 model. A similar result can 
be obtained for $B_{s}\to \mu^+ \mu^-$ using the corresponding bounds from $\Delta M_s$, which we have added in Figure \ref{bdmumu}. Also in this case, we find that 
there is not much room for a significant enhancement.
Investigating now also the implications for a suppression of these branching fractions, we find that these can be larger, but a very large effect here is also excluded.

\begin{figure}
\begin{center}
\includegraphics[height=5cm]{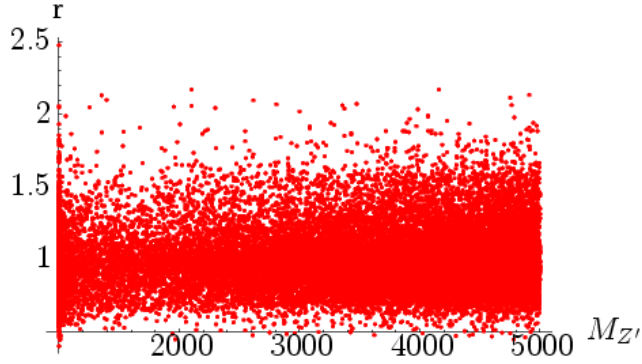}
\end{center}
\caption{\label{BmumuDMq} Deviation from unity of the factor $r$ introduced in the text, as depending on $M_{Z'}$.  }
\end{figure}

Finally, let us comment on the relation between $B_d \to \mu^+ \mu^-$ and $B_s \to \mu^+ \mu^-$ derived in \cite{Buras:2003td}. Here, one finds:
\begin{equation}
\frac{\mathrm{BR}(B_s \to \mu^+ \mu^-)}{\mathrm{BR}(B_d \to \mu^+ \mu^-)}=\frac{\hat B_d}{\hat B_s} \frac{\tau(B_s)}{\tau(B_d)} \frac{\Delta M_s}{\Delta M_d} r 
\end{equation}
This relation has the advantage that the form factors $F_{B_q}$ drop out, and that therefore the uncertainties are reduced significantly. It is valid with $r=1$ in the 
SM and any extension that has an MFV structure. In our model, however, we can expect significant departures from this relation, i.e. a value of $r$ that is not necessary
 unity. Exploring the possible violation of this relation, we are, of course interested in the range that $r$ can obtain. For this we scan over the entire allowed
parameter range to obtain all possible values of $r$. The result of this investigation is
shown in Figure \ref{BmumuDMq}, where the constraints from $\Delta M_d$, $\Delta M_s$ and $\sin 2 \beta$ are all included. We find that the SM relation can be broken by 
about $50\%$, with $r$ ranging from $r \approx 0.5-2$, while this range seems to be rather independent of the $Z'$ mass.
 It is clear that we could have expected these strong modifications, since the mass differences are significantly 
more sensitive to the new contributions due to the leptophobic structure. As a general conclusion to this section, we can therefore state just this: In the minimal 331 
model, we expect there to be stronger modifications in any quantity, in which leptons are not involved, i.e. in particular in the CP-violating asymmetries measuring
$\beta$ and $\beta_s$, as well as the mass differences.

%\begin{figure}
%\begin{center}
%\includegraphics[height=5cm]{Bsmumuscat.eps}
%\end{center}
%\caption{\label{bsmumu}Allowed range for the branching ratio $B_s \to \mu^+ \mu^-$ from $\Delta M_s$.}
%\end{figure}

\begin{figure}
\begin{center}
\includegraphics[height=6cm]{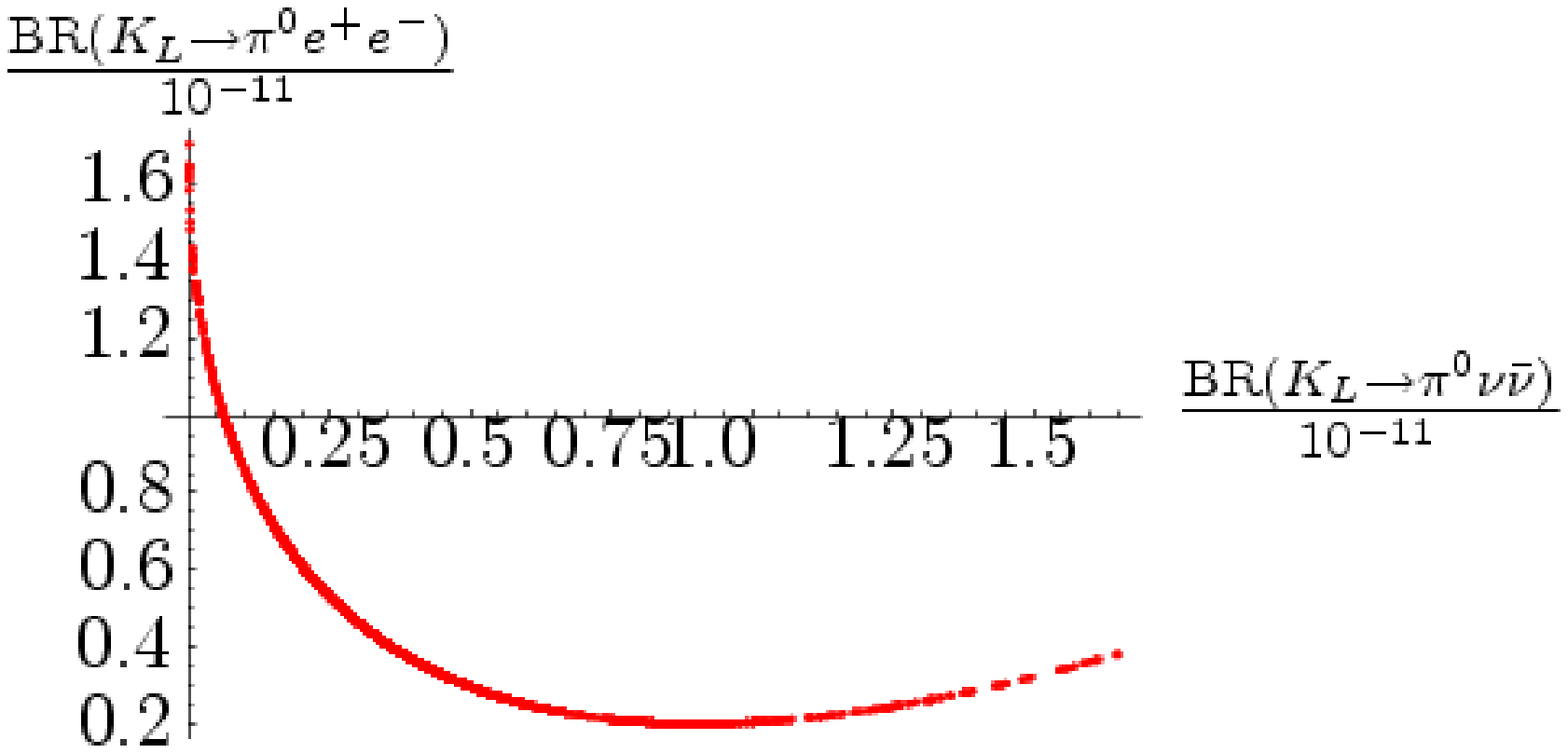}
\end{center}
\caption{\label{KeeKnunu}Contour in the $K_L \to \pi^0 e^+ e^-$-$\klpn$ plane.}
\end{figure}

\begin{figure}
\begin{center}
\includegraphics[height=6cm]{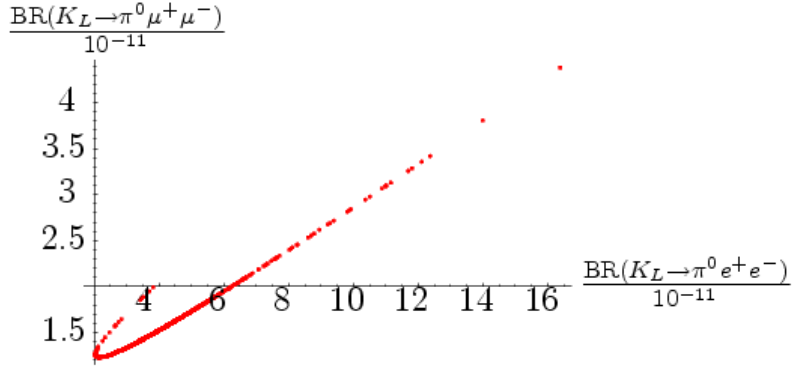}
\end{center}
\caption{\label{KeeKmumu}Analog to Fig. \ref{KeeKnunu} in the $K_L \to \pi^0 e^+ e^-$-$K_L \to \pi^0 \mu^+ \mu^-$ plane. 
A measurement of any two decays tests the minimal 331 model.}
\end{figure}

We conclude our numerical analysis with the compilation of Table \ref{Obs}, where we have collected the possible enhancements and suppressions in several observable 
quantities scanning the input parameters in a manner similar to the analysis of $r$ above.
We observe, that the value of $\sin 2 \beta$, as obtained from the combined $K \to  \pi \nu \bar \nu$ decays \cite{Buchalla:1994tr} can receive significant 
modifications, as well as both leptonic decays $K \to \pi^0 l^+ l^-$ which may also be rather strongly modified by the new contributions. Note, that in all these 
cases, there are, in particular, very strict lower bounds, valid for all of the $Z'$ mass range, that can not be circumvented. The stronger enhancement of the 
$K_L \to \pi^0 e^+ e^-$ branching fraction as compared to the one of $K_L \to \pi^0 \mu^+ \mu^-$ is a reflection of the fact that $\Delta y_{7V}$ is larger than 
$\Delta y_{7A}$ by a factor of 3. Also, a general feature of many models is that the decay $K_L \to \pi^0 e^+ e^-$ is subject to weaker modifications than the 
$\klpn$ decay, which is clearly
not the case in the minimal 331 model. We therefore show the contour in the observable plane of these two decays in Fig.~\ref{KeeKnunu}, 
which displays this feature rather nicely. Also, this 
contour allows an immediate test of the 331 model, if both decays are measured. The same is true also for the correlation of $K_L \to \pi^0 e^+ e^-$ and 
$K_L \to \pi^0 \mu^+ \mu^-$, which we add in Fig.~\ref{KeeKmumu}, in the spirit of \cite{Isidori:2004rb}.

\begin{table}[tb]
\center{
\begin{tabular}{|l|l|l|l|}
\hline
& Allowed Region for  & Allowed Region for  & Allowed Region for \\ 
& $M_{Z'}= 1~ \mathrm{TeV}$ & $M_{Z'}= 3~ \mathrm{TeV}$& $M_{Z'}= 5~ \mathrm{TeV}$ \\ \hline
$\beta|_{K \pi \nu \nu}$ & $(0- 45)^{\circ}$ & $(17-32)^{\circ}$ & $(21-29)^{\circ}$  \\
$\mathrm{BR}(K_L\to \pi^0 e^+e^-)/10^{-11}$ & $(2-17.5)$ & $(2.3-7.4)$ & $(3-5.5)$ \\
$\mathrm{BR}(K_L\to \pi^0 \mu^+\mu^-)/10^{-11}$ & $(1.2-4.6)$ & $(1.2-2.2)$ & $(1.3-1.8)$\\
\hline
\end{tabular}  
\caption{Ranges for several observable quantities affected by tree level $Z'$ exchange.}
\label{Obs}}
\end{table}

\section{Conclusions}\label{sec:conclusions}
We have analyzed in detail the flavor structure of the minimal 331 model, including for the first time explicitely the effects of new CP violating phases, as well as
the new data for $\Delta M_s$. This allowed us to analyze a larger set of observable quantities than has been done before. Here, we have concentrated on the contributions 
from the exchange of the new $Z'$ gauge boson, which transmits FCNC processes at tree level. We have used the experimentally measured
quantities $\Delta M_K$, $\varepsilon_K$, $\Delta M_{d/s}$ and $\sin 2 \beta |_{J/\psi K_S}$ to constrain the size of the new mixing matrix elements, depending on the 
mass of the $Z'$ boson. We have then taken these results to obtain bounds for several very clean rare decay processes, i.e. the decays $\kpn$, $\klpn$, $K_L \to \pi^0 l^+ l^-$ 
and $B_{d/s}\to \mu^+ \mu^- $. These upper bounds depend on the $Z'$ mass and can be used to exclude the minimal 331 model, or at least certain 
ranges of the $Z'$ mass. Let us summarize the results of the different steps in our analysis as follows
\begin{itemize}
\item FCNC processes are very well suited to constrain and explore the minimal 331 model, since the new contributions to EWP observables appear only at loop level, 
while the new FCNC effects appear already at tree level and are thus more significant. 
\item In the mixing sector of the neutral kaon system, we find that the imaginary part of $(\tilde V_{32}^* \tilde V_{31})^2$ is much stronger constrained than the real part.
Therefore, if we would like to saturate these bounds, we can take a purely real or imaginary  $(\tilde V_{32}^* \tilde V_{31})$. 
\item Concerning $\Delta M_{d/s}$, we find that modifications to both observables can take place as enhancements or suppressions in an equal manner, and that the 
measurements already significantly constrain the respective mixing matrix elements. We find, however that the bounds from $\sin 2 \beta$ are somewhat stronger than
those from $\Delta M_d$, depending on the relative phase of the new contributions. Additionally, the phase $\phi_s$, as measured in the mixing induced asymmetry of
$B^0_s \to J/\psi \phi$ can be large, since it is basically unconstrained, as of now. At the same time, the new contributions could solve a potential discrepancy 
between the measured values of $\sin 2 \beta$ and $|V_{ub}|$, in case the corresponding discrepancy persists.    
\item There are potentially significant modifications in both $K \to \pi \nu \bar \nu$ effects, depending on the phase structure of the new mixing matrix and, of course, 
the $Z'$ mass. In fact, measuring both decays allows one to unambiguously determine the new phase as well as the absolute value of the combination 
$\tilde V^*_{32} \tilde V_{31}/M_{Z'}^2$. The present experimental central value for $\mathrm{BR}(\kpn)$ can be reached, but only for rather low values of $M_{Z'}$.
Also, we point out, that there are two "branches" in the $\klpn-\kpn$ plane, similar to the signature in the Littlest Higgs model with T-parity, but the possible enhancements
are significantly smaller than in that model. On the other hand, the signature in $K_L \to \pi^0 e^+ e^-$ is stronger than in the LHT model, in particular in relation
to the enhancement of $\klpn$. This difference is due to the fact that vector and axial-vector contributions partially cancel out in the $V-A$ difference, to which 
$\klpn$ is sensitive, while the individually large modification of the vector component affects $K_L \to \pi^0 e^+ e^-$.
 
\item Next, we have then analyzed the impact of the bounds from $\Delta M_{d/s}$ on the decays $B_{d/s}\to \mu^+ \mu^-$. Here, we find that large enhancements seem 
impossible, while significant suppressions of both branching ratios can be obtained. 
\item Finally, we have briefly investigated some correlations and relations between several decays that hold in the SM, but are expected to be violated in 
the minimal 331 model, in particular, if new CP violating phases are present. For example, we find that the relation \cite{Buras:2003td} between $\Delta M_{d/s}$ and 
$\mathrm{BR}(B_{d/s}\to \mu^+ \mu^-)$ can be rather strongly violated by up to $50\%$.  
\item The most general conclusion to draw from this analysis is that, in general, we expect stronger modifications in those observables, that do not involve leptonic
couplings, since these are suppressed in comparison to the quark coupling. In this context, the phase $\beta_s$ as measured in the mixing induced asymmetry 
$B^0_s \to J/\psi \phi$ becomes extremely interesting.
\end{itemize}
Finally, we would like to point out that the minimal 331 model is only one example of a model with an additional $Z'$ boson, but has many features that any such model
should share, such as the correspondence of the bounds from $\Delta M_i$ to effects in the rare decays, which will stay the same in any such model, subject to small 
modifications from different lepton couplings. Here, we would again like to point out that the lepton coupling to the $Z'$ is suppressed in our model by a factor of
$\sqrt{(1-4 \sin \theta_W)}$, so that stronger effects should be expected in a general model. On the other hand, the illustrated patterns in the rare decay sector remain 
the same, in particular the possibility of obtaining information on phase structure and absolute values from measurements of both $K \to \pi \nu \bar \nu $ decays. 
The same is true for the correlation between $\Delta M_{d/s}$ and $\sin 2 \beta_{d/s}$, 
implicitly stated in Eqs. (\ref{DeltaMq}) and (\ref{Phidcorrection}). Therefore, our analysis of the minimal 331 model can also serve as an example-analysis of this more
general situation.

\noindent
{\bf Acknowledgments}\\
\noindent
We would like to thank A.J. Buras for valuable discussions and C. Hagedorn for several valuable comments on the manuscript. 
F.S. acknowledges financial support from the Deutsche Forschungsgemeinschaft (DFG) and from the 
``Bundesministerium f\"ur Bildung und Forschung (BMBF)" under contract 05HT6WOA.

\begin{appendix}

\section{Feynman Rules for Vertices}\label{sec:FR}
In this section, we list all the Feynman Rules relevant to our calculation. We define $P_L \equiv \frac{\gamma^{\mu}}{2} (1-\gamma^5)$ and 
$P_R \equiv \frac{\gamma^{\mu}}{2} (1+\gamma^5)$.
\indent

\subsubsection*{Quark - $Z'$ - vertices}
\indent

% uuZ' flavour changing
\begin{tabular}{ll}
\begin{picture}(150,70)(0,-10)
%left horizontal line
\Photon(10,0)(60,0){3}{4}
\Text(0,0)[c]{$Z'_{\mu}$}
%right horizontal line
\ArrowLine(60,0)(110,0)
\Text(120,0)[c]{$u_j$}
%upper vertical line
\ArrowLine(60,50)(60,0)
\Text(65,45)[l]{$u_i$}
%blob
\Vertex(60,0){2}
\end{picture}
&
\raisebox{35\unitlength}{
\begin{minipage}{5cm}
\lefteqn{
i \: \frac{g c_W}{\sqrt{3} \sqrt{1 - 4 s_W^2}} \: U_{3 U_i} U_{3 U_j}^\ast  \: P_L
\qquad i,j=1,2,3 \,, i \neq j}
\end{minipage}
}
\end{tabular}

% uuZ' flavour conserving
\begin{tabular}{ll}
\begin{picture}(150,70)(0,-10)
%left horizontal line
\Photon(10,0)(60,0){3}{4}
\Text(0,0)[c]{$Z'_{\mu}$}
%right horizontal line
\ArrowLine(60,0)(110,0)
\Text(120,0)[c]{$u_i$}
%upper vertical line
\ArrowLine(60,50)(60,0)
\Text(65,45)[l]{$u_i$}
%blob
\Vertex(60,0){2}
\end{picture}
&
\raisebox{35\unitlength}{
\begin{minipage}{5cm}
\lefteqn{
 \frac{-i g }{\sqrt{3} c_W \sqrt{1 - 4 s_W^2}} \: \left( (1/2- s_W^2-c_W^2 U_{3 i} U_{3 i}^\ast) P_L - \:2 s_W^2 P_R \right) }
\end{minipage}
}
\end{tabular}

% ttZ' flavour conserving
%\begin{tabular}{ll}
%\begin{picture}(150,70)(0,-10)
%left horizontal line
%\Photon(10,0)(60,0){3}{4}
%\Text(0,0)[c]{$Z'_{\mu}$}
%right horizontal line
%\ArrowLine(60,0)(110,0)
%\Text(120,0)[c]{$t$}
%upper vertical line
%\ArrowLine(60,50)(60,0)
%\Text(65,45)[l]{$t$}
%blob
%\Vertex(60,0){2}
%\end{picture}
%&
%\raisebox{35\unitlength}{
%\begin{minipage}{5cm}
%%\lefteqn{
%%- i \: (\frac{g (1-2 s_W^2)}{2 \sqrt{3} c_W \sqrt{1 - 4 s_W^2}}- \frac{g c_W}{\sqrt{3} \sqrt{1 - 4 s_W^2}} \: u_{3 3} u_{3 3}^\ast )\: P_L \\
%% +i \:\frac{2 g s_W^2}{\sqrt{3} c_W \sqrt{1 - 4 s_W^2}} \:P_R
%%}
%\lefteqn{
%\frac{-ig}{\sqrt{3} c_W \sqrt{1 - 4 s_W^2}} \left((1/2 -s_W^2-c_W^2 U_{3 3} U_{3 3}^\ast)P_L - 2 s_W^2 P_R \right)
%}
%\end{minipage}
%}
%\end{tabular}

% ddZ' flavour changing
\begin{tabular}{ll}
\begin{picture}(150,70)(0,-10)
%left horizontal line
\Photon(10,0)(60,0){3}{4}
\Text(0,0)[c]{$Z'_{\mu}$}
%right horizontal line
\ArrowLine(60,0)(110,0)
\Text(120,0)[c]{$d_j$}
%upper vertical line
\ArrowLine(60,50)(60,0)
\Text(65,45)[l]{$d_i$}
%blob
\Vertex(60,0){2}
\end{picture}
&
\raisebox{35\unitlength}{
\begin{minipage}{5cm}
\lefteqn{
i \: \frac{g c_W}{\sqrt{3} \sqrt{1 - 4 s_W^2}} \: \tilde V_{3 d_i} \tilde V_{3 d_j}^\ast \: P_L
\qquad i,j=1,2,3 \,, i \neq j}
\end{minipage}
}
\end{tabular}

% ddZ' flavour conserving
\begin{tabular}{ll}
\begin{picture}(150,70)(0,-10)
%left horizontal line
\Photon(10,0)(60,0){3}{4}
\Text(0,0)[c]{$Z'_{\mu}$}
%right horizontal line
\ArrowLine(60,0)(110,0)
\Text(120,0)[c]{$d_i$}
%upper vertical line
\ArrowLine(60,50)(60,0)
\Text(65,45)[l]{$d_i$}
%blob
\Vertex(60,0){2}
\end{picture}
&
\raisebox{35\unitlength}{
\begin{minipage}{5cm}
\lefteqn{
\frac{-i g}{\sqrt{3} c_W \sqrt{1 - 4 s_W^2}} \left( (1/2- s_W^2 - c_W^2 \tilde V_{3i} \tilde V_{3i}^*) \: P_L + s_W^2\:P_R \right)}
\end{minipage}
}
\end{tabular}

% bbZ' flavour conserving
%\begin{tabular}{ll}
%\begin{picture}(150,70)(0,-10)
%left horizontal line
%\Photon(10,0)(60,0){3}{4}
%\Text(0,0)[c]{$Z'_{\mu}$}
%right horizontal line
%\ArrowLine(60,0)(110,0)
%\Text(120,0)[c]{$b$}
%upper vertical line
%\ArrowLine(60,50)(60,0)
%\Text(65,45)[l]{$b$}
%blob
%\Vertex(60,0){2}
%\end{picture}
%&
%\raisebox{35\unitlength}{
%\begin{minipage}{5cm}
%\lefteqn{
%-i \: (\frac{g (1-2 s_W^2)}{2\sqrt{3} c_W \sqrt{1 - 4 s_W^2}} - \frac{g c_W}{\sqrt{3} \sqrt{1 - 4 s_W^2}} \: v_{3 d_i} v_{3 d_j}^\ast) \: P_L -i \:\frac{g s_W^2}{\sqrt{3} c_W \sqrt{1 - 4 s_W^2}} \:P_R
%}
%\lefteqn{ \frac{-ig}{\sqrt{3} c_W \sqrt{1 - 4 s_W^2}} \left(( 1/2 -s_W^2 -c_W^2 \tilde V_{3 3} \tilde V_{3 3}^\ast)P_L+s_W^2 P_R \right)
%}
%\end{minipage}
%}
%\end{tabular}

% TTZ'
\begin{tabular}{ll}
\begin{picture}(150,70)(0,-10)
%left horizontal line
\Photon(10,0)(60,0){3}{4}
\Text(0,0)[c]{$Z'_{\mu}$}
%right horizontal line
\ArrowLine(60,0)(110,0)
\Text(120,0)[c]{$T$}
%upper vertical line
\ArrowLine(60,50)(60,0)
\Text(65,45)[l]{$T$}
%blob
\Vertex(60,0){2}
\end{picture}
&
\raisebox{35\unitlength}{
\begin{minipage}{5cm}
\lefteqn{
-i \: \frac{g}{\sqrt{3} c_W \sqrt{1 - 4 s_W^2}} \:  \left( (1-6 s_W^2) P_L - 5 s_W^2 P_R \right)
}
\end{minipage}
}
\end{tabular}

% DDZ'
\begin{tabular}{ll}
\begin{picture}(150,70)(0,-10)
%left horizontal line
\Photon(10,0)(60,0){3}{4}
\Text(0,0)[c]{$Z'_{\mu}$}
%right horizontal line
\ArrowLine(60,0)(110,0)
\Text(120,0)[c]{$D_i$}
%upper vertical line
\ArrowLine(60,50)(60,0)
\Text(65,45)[l]{$D_i$}
%blob
\Vertex(60,0){2}
\end{picture}
&
\raisebox{35\unitlength}{
\begin{minipage}{5cm}
\lefteqn{
i \: \frac{g}{\sqrt{3} c_W \sqrt{1 - 4 s_W^2}}  \: \left( (1-5 s_W^2) P_L + 4 s_W^2 P_R \right) \qquad i = 1,2
}
\end{minipage}
}
\end{tabular}

\subsubsection*{Lepton - $Y^\pm$ - vertices}
\indent

%\nul^CV
\begin{tabular}{ll}
\begin{picture}(150,70)(0,-10)
%left horizontal line
\Photon(10,0)(60,0){3}{4}
\Text(0,0)[c]{$Y_{\mu}$}
%right horizontal line
\ArrowLine(60,0)(110,0)
\Text(120,0)[c]{$l^C$}
%upper vertical line
\ArrowLine(60,50)(60,0)
\Text(65,45)[l]{$\nu_l$}
%blob
\Vertex(60,0){2}
\end{picture}
&
\raisebox{35\unitlength}{
\begin{minipage}{5cm}
\lefteqn{
i \: \frac{g}{\sqrt{2}} \: P_L
}
\end{minipage}
}
\end{tabular}

\subsubsection*{Lepton - $Y^{\pm\pm}$ - vertices}
\indent

%ll^CY
\begin{tabular}{ll}
\begin{picture}(150,70)(0,-10)
%left horizontal line
\Photon(10,0)(60,0){3}{4}
\Text(0,0)[c]{$Y_{\mu}$}
%right horizontal line
\ArrowLine(60,0)(110,0)
\Text(120,0)[c]{$l^C$}
%upper vertical line
\ArrowLine(60,50)(60,0)
\Text(65,45)[l]{$l$}
%blob
\Vertex(60,0){2}
\end{picture}
&
\raisebox{35\unitlength}{
\begin{minipage}{5cm}
\lefteqn{
-i \: \frac{g}{\sqrt{2}} \: P_L
}
\end{minipage}
}
\end{tabular}

\end{appendix}

\end{document}